\documentclass[prb,amsmath,superscriptaddress,showpacs,showkeys,twocolumn]{revtex4}
\usepackage{amssymb,amsmath}   
\usepackage[dvips]{graphicx}   
\usepackage{verbatim}   
\usepackage{color}      
\usepackage{subfigure}  
\usepackage{hyperref}   
\usepackage{gensymb}
\usepackage{epstopdf}
\usepackage{natbib}

\begin{document}

\def\tauv{\mbox{\boldmath{$\tau$}}}
\def\qv{\mbox{\boldmath{$q$}}}
\def\Rv{\mbox{\boldmath{$R$}}}
\def\hv{\mbox{\boldmath{$h$}}}
\def\mv{\mbox{\boldmath{$m$}}}
\def\Sv{\mbox{\boldmath{$S$}}}
\def\Gv{\mbox{\boldmath{$G$}}}
\def\Pv{\mbox{\boldmath{$P$}}}

\title{Landau theory for the phase diagram of the multiferroic $\textbf{Mn}_\textbf{{1-x}}\textbf{(Fe,Zn,Mg)}_\textbf{{x}}\textbf{WO}_\textbf{{4}}$}
\author{Shlomi Matityahu}
\affiliation{Department of Physics, Ben-Gurion University, Beer
Sheva 84105, Israel}
\author{Amnon Aharony}
\email{aaharony@bgu.ac.il} \altaffiliation{Also at Tel Aviv
University.} \affiliation{Department of Physics, Ben-Gurion
University, Beer Sheva 84105, Israel} \affiliation{Ilse Katz
Center for Meso- and Nano-Scale Science and Technology, Ben-Gurion
University, Beer Sheva 84105, Israel}
\author{Ora Entin-Wohlman}
\altaffiliation{Also at Tel Aviv University.}
\affiliation{Department of Physics, Ben-Gurion University, Beer
Sheva 84105, Israel} \affiliation{Ilse Katz Center for Meso- and
Nano-Scale Science and Technology, Ben-Gurion University, Beer
Sheva 84105, Israel}
\date{\today}
\begin{abstract}
We present a theoretical analysis of the temperature-magnetic
field-concentration phase diagram of the multiferroic
Mn$^{}_{1-x}$M$^{}_{x}$WO$^{}_{4}$ (M=Fe, Zn, Mg), which exhibits
three ordered phases, with collinear and non-collinear
incommensurate and with a commensurate magnetic order. The middle
phase is also ferroelectric. The analysis uses a
semi-phenomenological Landau theory, based on a Heisenberg
Hamiltonian with a single-ion anisotropy. With a small number of
adjustable parameters, the Landau theory gives an excellent fit to
all three transition lines, as well as the magnetic and the
ferroelectric order parameters. The fit of the magnetic and
ferroelectric order parameters is further improved by including
the effect of fluctuations near the transitions. We demonstrate
the highly frustrated nature of these materials and suggest a
simple explanation for the dramatic effects of doping with
different magnetic ions at the Mn sites. The model enables an
examination of different sets of exchange couplings that were
proposed by a number of groups. Small discrepancies are probably a
consequence of small errors in the experimental magnetic
parameters. In addition, using the Ginzburg criterion we estimate
the temperature range in which fluctuations of the order
parameters become important.
\end{abstract}

\maketitle
\section{Introduction}
\label{Intro} Type \textrm{II} Magnetoelectric multiferroics are
materials which exhibit coexistence between certain types of
long-range magnetic order and a ferroelectric order. These
materials are usually characterized by a strong magnetoelectric
coupling between their electric and magnetic degrees of freedom.
The magnetoelectric effect enables the control of the electric
polarization by a magnetic field, or the control of the
magnetization by an electric field. The study of magnetoelectric
multiferroics is thus of great interest in condensed matter
physics, both from basic research and technological applications
points of view.\cite{KD09,CSW07,FM05,KT07} In recent years, the
interest in this field has grown after the discovery of new
materials with a large magnetoelectric effect, such as
TbMnO$^{}_{3}$,\cite{KT03} TbMn$^{}_{2}$O$^{}_{5}$,\cite{HN04}
Ni$^{}_{3}$V$^{}_{2}$O$^{}_{8}$,\cite{LG05}
CuFeO$^{}_{2}$,\cite{KT06} and CoCr$^{}_{2}$O$^{}_{4}$.\cite{YY06}
In those oxides, ferroelectricity appears in conjunction with a
noncollinear spiral magnetic phase, which breaks spatial inversion
symmetry, and therefore allows the appearance of an electric
polarization.

There are two different approaches to the theoretical treatment of
such noncollinear magnetoelectric multiferroics. One approach is
based on first principles calculations using the density
functional theory (DFT).\cite{PS09} The second approach constructs
a model Hamiltonian dictated by symmetry
considerations.\cite{LG05,MM06,HAB07} Different mechanisms for the
magnetoelectric coupling can then be
suggested.\cite{HAB06,KD09,KH05,SI06} In this paper we develop a
semi-phenomenological model for describing the magnetic phase
transitions of Mn$^{}_{1-x}$M$^{}_{x}$WO$^{}_{4}$ (M=Fe, Zn, Mg)
and the induced ferroelectric polarization. The model is
semi-phenomenological in the sense that some of the parameters can
be deduced from existing experimental data, while the others are
purely phenomenological. The multiferroic MnWO$^{}_{4}$ is a
natural choice for such an approach, due to the vast experimental
data that exists in the literature.

MnWO$^{}_{4}$ crystallizes in the wolframite structure, which
belongs to the monoclinic space group P2/c with
$\beta\approx91\degree$. The unit cell includes two magnetic
Mn$^{2+}$ ions with spin $S=5/2$ and orbital angular momentum
$L=0$ at positions $\tauv^{}_1=(0.5,y,0.25)$ and
$\tauv^{}_2=(0.5,1-y,0.75)$ (in units of the primitive lattice
vectors) with $y=0.685$.\cite{LG93} In zero magnetic field,
MnWO$^{}_{4}$ undergoes three successive phase transitions at
temperatures $T^{}_{N3}\approx13.5K$, $T^{}_{N2}\approx12.3-12.7K$
and $T^{}_{N1}\approx7-8K$ to phases which are called AF3, AF2,
and AF1, respectively.\cite{LG93,AAH06,TK06} According to neutron
diffraction experiments,\cite{LG93} AF3 is an incommensurate (IC)
antiferromagnetic phase with a collinear sinusoidal structure, AF2
is an incommensurate antiferromagnetic phase with an
elliptical-spiral structure, and AF1 is a commensurate (C)
antiferromagnetic phase with a collinear
$\uparrow\uparrow\downarrow\downarrow$ structure. The propagation
vectors are $\qv^{}_{IC}=(-0.214,0.5,0.457)$ (in units of the
primitive reciprocal lattice vectors) for AF2 and AF3, and
$\qv^{}_{C1,2}=(\pm0.25,0.5,0.5)$ for AF1. In AF3 and AF1, the
magnetic moments of the Mn$^{2+}$ ions align along the easy axis
of magnetization, which lies in the $ac$-plane and forms an angle
of $\approx35\degree-37\degree$ with the $a$ axis. Different
studies\cite{AAH06,TK06} reveal that a ferroelectric polarization,
which is oriented along the $b$ axis, develops in the AF2 phase.

As opposed to MnWO$^{}_{4}$, other isomorphic wolframite
structures like FeWO$^{}_{4}$, CoWO$^{}_{4}$ and NiWO$^{}_{4}$
show only a single magnetic phase transition to a simple
commensurate antiferromagnetic phase with the propagation vector
$\qv^{}_{}=(0.5,0,0)$.\cite{WH77} Those observations suggest that
unlike the isomorphic structures, MnWO$^{}_{4}$ constitutes a
highly frustrated system with complex competing interactions. The
competition between the different interactions manifests itself in
the sensitivity of the phase diagram to doping with different
transition metal ions at the Mn sites. It turns out that a small
Fe concentration suppresses the ferroelectric phase AF2 and
expands the stabilization range of AF3 and
AF1.\cite{CRP09,YF08,CRP08} In contrast to Fe doping, it has been
reported\cite{SYS09} that a small Co concentration stabilizes the
ferroelectric phase at the expense of the AF1 phase. A
quantitative and microscopic understanding of the effect of Fe and
Co doping on the multiferroic properties and the phase diagram of
MnWO$^{}_{4}$ is quite complicated, since the exchange couplings
of the M-M and M-Mn (M=Fe, Co) interactions, as well as the
anisotropy parameters are not known. In order to overcome some of
these problems, a much simpler magnetic system has been achieved
by the partial substitution of Mn ions by the non-magnetic ions
Zn$^{2+}$ and Mg$^{2+}$.\cite{CRP11,ML09} Those studies reveal
that the AF1 phase is strongly suppressed as a result of magnetic
ions dilution by non-magnetic substituents.

The frustrated nature of MnWO$^{}_{4}$ was demonstrated by
Ehrenberg \textit{et al.}.\cite{EH99} Using inelastic neutron
scattering they extracted 9 exchange couplings $J^{}_1-J^{}_9$ for
the superexchange interactions among the Mn ions. Later, Tian
\textit{et al.}\cite{TC09} proposed different values for the 9
exchange couplings based on DFT calculations. Those values depend
on an unknown on-site repulsion energy. Moreover, the authors have
noted that generally DFT calculations tend to overestimate the
magnitude of exchange interactions.\cite{TC09} Recently, the
experimental data have been expanded.\cite{YF11} In that study, Ye
\textit{et al.} suggested some corrections for the values of the
exchange couplings, and included two additional ones, $J^{}_{10}$
and $J^{}_{11}$. The two sets of experimental exchange couplings
are summarized in Table ~\ref{tab:Exchange couplings}. The model
we describe may help to compare these different sets of exchange
couplings, by examining their consistency with different
experimental observations.
\begin{table*} \tiny
\centering \caption{\label{tab:Exchange couplings} Superexchange
couplings for the Mn$^{2+}$ ion at $\tauv^{}_1=(0.5,y,0.25)$
according to different inelastic neutron scattering studies. We
denote $z=1-y$, $w=2-y$ and $u=1+y$. The values are presented in
units of $k^{}_{B}K$.\cite{comment1}}
    \begin{tabular}{c|c|c|c|c|c|c|c|c|c|c|c|c}
    \hline
    \hline
     & $J^{}_1$ & $J^{}_2$ & $J^{}_3$ & $J^{}_4$ & $J^{}_5$ & $J^{}_6$ & $J^{}_7$ & $J^{}_8$ & $J^{}_9$ & $J^{}_{10}$ & $J^{}_{11}$ & D \\ \hline
    Neighbors & $\left(\frac{1}{2},z,\frac{3}{4}\right)$ & $\left(\frac{1}{2},w,\frac{3}{4}\right)$ & $\left(\frac{3}{2},y,\frac{1}{4}\right)$ & $\left(\frac{1}{2},y,\frac{5}{4}\right)$ & $\left(\frac{1}{2},u,\frac{1}{4}\right)$ & $\left(\frac{3}{2},z,\frac{3}{4}\right)$ & $\left(-\frac{1}{2},z,\frac{3}{4}\right)$ & $\left(\frac{3}{2},w,\frac{3}{4}\right)$ & $\left(-\frac{1}{2},w,\frac{3}{4}\right)$ & $\left(\frac{3}{2},y,\frac{5}{4}\right)$ & $\left(\frac{3}{2},y,-\frac{3}{4}\right)$ & \\
                      & $\left(\frac{1}{2},z,-\frac{1}{4}\right)$ & $\left(\frac{1}{2},w,-\frac{1}{4}\right)$ & $\left(-\frac{1}{2},y,\frac{1}{4}\right)$ & $\left(\frac{1}{2},y,-\frac{3}{4}\right)$ & $\left(\frac{1}{2},-z,\frac{1}{4}\right)$ & $\left(-\frac{1}{2},z,-\frac{1}{4}\right)$ & $\left(\frac{3}{2},z,-\frac{1}{4}\right)$ & $\left(\frac{3}{2},w,-\frac{1}{4}\right)$ & $\left(\frac{3}{2},w,-\frac{1}{4}\right)$ & $\left(-\frac{1}{2},y,-\frac{3}{4}\right)$ & $\left(-\frac{1}{2},y,\frac{5}{4}\right)$ & \\
    Ref. \onlinecite{EH99} & -0.195 & -0.135 & -0.423 &0.414 & 0.021 & -0.509 & 0.023 & 0.491 & -1.273 & - & - & 0.568 \\
    Ref. \onlinecite{YF11} & -1.95(1) & -0.18(1) & -1.48(1) & -1.21(1) & 0.23(1) & -1.99(1) & -0.56(1) & 0.09(1) & -1.21(1) & -0.7(1) & 0.09(1) & 0.84(1) \\ \hline \hline
    \end{tabular}
\end{table*}

The outline of the paper is as follows: in Sec. \ref{Sec 1} we
define the model. In Sec. \ref{Sec 2} the results of the model are
derived. In Sec. \ref{Sec 3} the model parameters are fitted by
comparing its results with different experimental observations.
Here we compare the two sets of experimental exchange couplings
with the fitted parameters. In Sec. \ref{Sec 4} the Ginzburg
criterion is applied to the specific case of the multiferroic
MnWO$^{}_{4}$, in order to examine whether the mean-field theory
approach is valid. We conclude in Sec. \ref{Sec 5} with a brief
summary.
\section{The model}
\label{Sec 1} In this section we develop the semi-phenomenological
model. The spin Hamiltonian consists of a Heisenberg term with a
single-ion anisotropy, which favors an easy axis in the
$ac$-plane. According to experiments, the spin component along the
hard axis in the $ac$-plane does not order in any of the phases.
Furthermore, the transitions are almost not influenced by an
external magnetic field along the hard axis. Hence we omit the
hard axis component from the calculations and write the spin as
$\Sv(\Rv+\tauv)=S^{}_{x}(\Rv+\tauv)\hat{\bf
x}+S^{}_{b}(\Rv+\tauv)\hat{\bf b}$, where $x$ denotes the easy
axis in the $ac$-plane and $b$ denotes the axis perpendicular to
the $ac$-plane. Here $\Sv(\Rv+\tauv)$ is the thermal average of
the dimensionless classical spin at position $\Rv+\tauv$, where
$\Rv$ is a lattice vector and $\tauv$ is one of the two basis
vectors $\tauv^{}_{1}$, $\tauv^{}_{2}$ in the unit cell,
indicating the locations of the Mn$^{2+}$ ions. We study the
following Hamiltonian:
\begin{widetext}
\begin{align}
\label{eq:Hamiltonian}
&H^{}_{mag}=-\frac{1}{2}\sum_{\Rv,\Rv'}\sum_{\tauv,\tauv'=\tauv^{}_{1},\tauv^{}_{2}}J(\Rv+\tauv,\Rv'+\tauv')
\Sv(\Rv+\tauv)\cdot\Sv(\Rv'+\tauv')-\frac{1}{2}D\sum_{\Rv}\sum_{\tauv=\tauv^{}_{1},\tauv^{}_{2}}{S^{}_{x}}^2(\Rv+\tauv).
\end{align}
\end{widetext}
Here $J(\Rv+\tauv,\Rv'+\tauv')$ is the superexchange interaction
energy which couples the spins at $\Rv+\tauv$ and $\Rv'+\tauv'$,
and $D$ is a positive single-ion anisotropy energy. To find an
expression for the magnetic free energy of the system, we expand
the entropy in the spin components up to the fourth order
\begin{align}
\label{eq:Entropy}
T\texttt{S}&=-\frac{1}{2}aT\sum_{\Rv}\sum_{\tauv=\tauv^{}_{1},\tauv^{}_{2}}{\Sv}^2(\Rv+\tauv)\nonumber\\
&-b\sum_{\Rv}\sum_{\tauv=\tauv^{}_{1},\tauv^{}_{2}}{\Sv}^4(\Rv+\tauv),
\end{align}
where $a$ and $b$ are positive parameters, and $T$ is the
temperature. Equation \eqref{eq:Entropy} gives the entropy
relative to the high temperature paramagnetic phase (denoted by P)
and thus the expression is negative. Combining Eqs.
\eqref{eq:Hamiltonian} and \eqref{eq:Entropy} we obtain the
magnetic free energy
\begin{widetext}
\begin{align}
&F^{}_{mag}=\frac{1}{2}\sum_{\Rv,\Rv'}\sum_{\tauv,\tauv'=\tauv^{}_{1},\tauv^{}_{2}}\sum_{\alpha,\beta=1}^2\chi^{-1}_{\alpha\beta}
(\Rv+\tauv,\Rv'+\tauv')S^{}_{\alpha}(\Rv+\tauv)
S^{}_{\beta}(\Rv'+\tauv')+b\sum_{\Rv}\sum_{\tauv=\tauv^{}_{1},\tauv^{}_{2}}{\Sv}^4(\Rv+\tauv),
\end{align}
\end{widetext}
where the $4\times4$ inverse susceptibility matrix is block
diagonal
\begin{align}
\chi^{-1}_{\alpha\beta}(\Rv+\tauv,\Rv'+\tauv')&=\big[(aT-D^{}_{\alpha})\delta^{}_{\Rv,\Rv'}\delta^{}_{\tau,\tau'}\nonumber\\
&-J(\Rv+\tauv,\Rv'+\tauv')\big]\delta^{}_{\alpha,\beta},
\end{align}
with $D^{}_{1}=D^{}_{x}=D$ and $D^{}_{2}=D^{}_{b}=0$. Below, we
exploit the Fourier transforms of the spin components,
\label{eq:Fourier transform}
\begin{align}
&S^{}_{\alpha}(\qv,\tauv)=\frac{1}{N}\sum_{\Rv}S^{}_{\alpha}(\Rv+\tauv)e^{i\qv\cdot(\Rv+\tauv)},\nonumber\\
&S^{}_{\alpha}(\Rv+\tauv)=\sum_{\qv}S^{}_{\alpha}(\qv,\tauv)e^{-i\qv\cdot(\Rv+\tauv)}.
\end{align}
Here $\qv$ is in the first Brillouin zone and $N$ is the number of
unit cells. In terms of the Fourier transform, the magnetic free
energy per unit cell, $f^{}_{mag}\equiv{F^{}_{mag}/N}$, is:
\begin{widetext}
\begin{align}
\label{eq:magnetic_free_energy}
f^{}_{mag}&=\frac{1}{2}\sum_{\tauv,\tauv'=\tauv^{}_{1},\tauv^{}_{2}}\sum_{\alpha,\beta=1}^2\sum_{\qv}\chi^{-1}_{\alpha\beta}(\qv;\tauv,\tauv')
{S^{}_{\alpha}}^\ast(\qv,\tauv)S^{}_{\beta}(\qv,\tauv')\nonumber\\
&+b\sum_{\Gv}\sum_{\tauv=\tauv^{}_{1},\tauv^{}_{2}}\sum_{\qv^{}_1,\qv^{}_2,\qv^{}_3,\qv^{}_4}e^{-i\Gv\cdot\tauv}\big[S^{}_{x}(\qv^{}_1,\tauv)S^{}_{x}(\qv^{}_2,\tauv)S^{}_{x}(\qv^{}_3,\tauv)S^{}_{x}(\qv^{}_4,\tauv)\nonumber\\
&+S^{}_{b}(\qv^{}_1,\tauv)S^{}_{b}(\qv^{}_2,\tauv)S^{}_{b}(\qv^{}_3,\tauv)S^{}_{b}(\qv^{}_4,\tauv)
+2S^{}_{x}(\qv^{}_1,\tauv)S^{}_{x}(\qv^{}_2,\tauv)S^{}_{b}(\qv^{}_3,\tauv)S^{}_{b}(\qv^{}_4,\tauv)\big]\delta(\qv^{}_1+\qv^{}_2+\qv^{}_3+\qv^{}_4-\Gv),
\end{align}
\end{widetext}
where $\Gv$ is a reciprocal lattice vector and the Fourier
transform of the inverse susceptibility matrix is given by the
block diagonal hermitian matrix
\begin{align}
\label{eq:chi_matrix}
\chi^{-1}_{\alpha\beta}(\qv;\tauv,\tauv')&=\big[(aT-D^{}_{\alpha})\delta^{}_{\tauv,\tauv'}\nonumber\\
&-J(\qv;\tauv,\tauv')\big]\delta^{}_{\alpha,\beta},
\end{align}
with $J(\qv;\tauv,\tauv')$ being the Fourier transform of the
$2\times2$ matrix $J(\Rv+\tauv,\Rv'+\tauv')$
\begin{align}
\label{eq:J_matrix} J(\qv;\tauv,\tauv')=
\sum_{\Rv}J(\tauv,\Rv+\tauv')e^{-i\qv\cdot(\Rv+\tauv'-\tauv)}.
\end{align}
In the last expression the sum is over all lattice vectors $\Rv$.
The four eigenvalues of the matrix \eqref{eq:chi_matrix} are
\begin{align}
\label{eq:eigenvalues}
&\zeta^{}_{\pm,x}(\qv,T)=aT-D-\lambda^{}_{\pm}(\qv), \nonumber\\
&\zeta^{}_{\pm,b}(\qv,T)=aT-\lambda^{}_{\pm}(\qv),
\end{align}
and the corresponding eigenvectors are
\begin{align}
\label{subeq:eigenvectors}
\Sv^{}_{\pm,x}(\qv)=\frac{1}{\sqrt{2}}\begin{pmatrix} 1\\ \pm{e}^{-i\phi(\qv)}\\ 0\\ 0 \end{pmatrix}, \nonumber\\
\Sv^{}_{\pm,b}(\qv)=\frac{1}{\sqrt{2}}\begin{pmatrix} 0\\ 0\\
1\\ \pm{e}^{-i\phi(\qv)} \end{pmatrix}.
\end{align}
Here, $\lambda^{}_{\pm}$ are the two eigenvalues of the matrix
\eqref{eq:J_matrix} and $\phi(\qv)$ is the phase of $J(\qv;1,2)$.
Assuming 11 exchange couplings as in Ref. \onlinecite{YF11}, these
two eigenvalues are given by
\begin{align}
\label{eq:lambda} \lambda^{}_{\pm}(\qv)&=\pm
2\sqrt{{\Lambda^{}_{2}}^2(\qv)+{\Lambda^{}_{3}}^2(\qv)+2\cos(2\pi
q_{b})\Lambda^{}_{2}(\qv)\Lambda^{}_{3}(\qv)}\nonumber\\
&+2\Lambda^{}_{1}(\qv),
\end{align}
with the following definitions:
\begin{align}
\Lambda^{}_{1}(\qv)&=J^{}_{3}\cos(2\pi q^{}_{a})+J^{}_{4}\cos(2\pi q^{}_{c})\nonumber\\
&+J^{}_{5}\cos(2\pi q^{}_{b})+J^{}_{10}\cos\big[2\pi(q^{}_{a}+q^{}_{c})\big]\nonumber\\
&+J^{}_{11}\cos\big[2\pi(q^{}_{a}-q^{}_{c})\big], \nonumber\\
\Lambda^{}_{2}(\qv)&=J^{}_{1}\cos(\pi
q^{}_{c})+J^{}_{6}\cos\big[2\pi(q^{}_{a}+\frac{q^{}_{c}}{2})\big]\nonumber\\
&+J^{}_{7}\cos\big[2\pi(q^{}_{a}-\frac{q^{}_{c}}{2})\big], \nonumber\\
\Lambda^{}_{3}(\qv)&=J^{}_{2}\cos(\pi
q^{}_{c})+J^{}_{8}\cos\big[2\pi(q^{}_{a}+\frac{q^{}_{c}}{2})\big]\nonumber\\
&+J^{}_{9}\cos\big[2\pi(q^{}_{a}-\frac{q^{}_{c}}{2})\big].
\end{align}
Now let us transform to magnetic normal coordinates
\begin{align}
\begin{pmatrix} S^{}_x(\qv,1) \\ S^{}_x(\qv,2) \\
S^{}_b(\qv,1) \\ S^{}_b(\qv,2)
\end{pmatrix}&=\sigma^{}_{+,x}(\qv)\Sv^{}_{+,x}(\qv)+\sigma^{}_{-,x}(\qv)\Sv^{}_{-,x}(\qv)\nonumber\\
&+\sigma^{}_{+,b}(\qv)\Sv^{}_{+,b}(\qv)+\sigma^{}_{-,b}(\qv)\Sv^{}_{-,b}(\qv).
\end{align}
Here, $\sigma^{}_{+,x}(\qv)$, $\sigma^{}_{-,x}(\qv)$,
$\sigma^{}_{+,b}(\qv)$ and $\sigma^{}_{-,b}(\qv)$ are the magnetic
order parameters for a magnetic structure with wave vector $\qv$.
The diagonal form of the magnetic free energy
\eqref{eq:magnetic_free_energy} is therefore
\begin{align}
f^{}_{mag}&=\frac{1}{2}\sum_{\qv}\big[\zeta^{}_{+,x}(\qv,T)\left|\sigma^{}_{+,x}(\qv)\right|^2+
\zeta^{}_{-,x}(\qv,T)\left|\sigma^{}_{-,x}(\qv)\right|^2 \nonumber\\
&+\zeta^{}_{+,b}(\qv,T)\left|\sigma^{}_{+,b}(\qv)\right|^2+
\zeta^{}_{-,b}(\qv,T)\left|\sigma^{}_{-,b}(\qv)\right|^2\big]\nonumber\\
&+O({\sigma}^{4}).
\end{align}
At high enough temperatures, the eigenvalues
\eqref{eq:eigenvalues} are all positive and therefore the stable
phase is the paramagnetic one. As we lower the temperature, we
reach a critical temperature for which one of the eigenvalues
vanishes. We denote the wave vector for which one of the
eigenvalues vanishes first as $\qv^{}_{IC}$. Since
$\lambda^{}_{+}(\qv)>\lambda^{}_{-}(\qv)$ and $D>0$, the first
eigenvalue which reaches zero is $\zeta^{}_{+,x}$. At the
temperature $T^{(0)}_{N3}$ at which $\zeta^{}_{+,x}=0$ there is a
phase transition from the paramagnetic phase to the AF3 phase, in
which $\sigma^{}_{+,x}(\qv^{}_{IC})\neq0$ but all other order
parameters remain zero. At the second transition
AF3$\rightarrow$AF2, the order parameter
$\sigma^{}_{+,b}(\qv^{}_{IC})$ orders as well. This is true
provided that
\begin{equation}
\label{eq:condition}
\lambda^{}_{+}(\qv^{}_{IC})-\lambda^{}_{-}(\qv^{}_{IC})>D.
\end{equation}
The last condition ensures that $\zeta^{}_{+,b}(\qv^{}_{IC},T)$
vanishes before $\zeta^{}_{-,x}(\qv^{}_{IC},T)$ as the temperature
is lowered. Henceforth, we will omit the plus sign in the order
parameters subscript.

To describe the electric polarization, we need to add an electric
free energy and a magnetoelectric coupling term to the magnetic
free energy. Assuming a homogeneous polarization, the expression
for the electric free energy to lowest order is
\begin{equation}
f^{}_{el}=V^{}_{cell}\sum_{\alpha=1}^3\frac{{P^{}_{\alpha}}^2}{2\chi^{0}_{E,\alpha}},
\end{equation}
where $V^{}_{cell}$ is the volume of the unit cell, $\Pv$ is the
ferroelectric order parameter and $\chi^{0}_{E,\alpha}$ is the
high-temperature electric susceptibility along the $\alpha$
direction. By symmetry considerations,\cite{HAB07} the allowed
magnetoelectric coupling term of the lowest order in the
incommensurate phases is
\begin{equation}
f^{}_{int}=r\left|\sigma^{}_{x}(\qv^{}_{IC})\right|\left|\sigma^{}_{b}(\qv^{}_{IC})\right|\sin(\varphi^{}_{x}-\varphi^{}_{b})P^{}_{b},
\end{equation}
where $\varphi^{}_{x}$ and $\varphi^{}_{b}$ are the phases of
$\sigma^{}_{x}(\qv^{}_{IC})$ and $\sigma^{}_{b}(\qv^{}_{IC})$,
respectively, and $r$ is a small real magnetoelectric coupling
parameter. Below we examine the results of the model.

\section{Phase boundaries and order parameters}
\label{Sec 2}
\subsection{\textbf{MnWO}$^{}_{\textbf{4}}$ without magnetic fields}
The wave vector $\qv^{}_{IC}$ that characterizes the AF3 and AF2
phases is determined by maximizing the eigenvalue
$\lambda^{}_{+}(\qv)$ for a given set of coupling energies
$\{J^{}_i\}$. After carrying out the maximization procedure, we
can find the first transition temperature by equating
$\zeta^{}_{+,x}$ to zero for $\qv=\qv^{}_{IC}$:
\begin{align}
\label{eq:T_N3}
&T^{(0)}_{N3}=\frac{\lambda^{}_{+}(\qv_{IC})+D}{a}.
\end{align}
The index 0 indicates that this is the transition temperature in
the absence of external magnetic fields. By transforming to normal
magnetic coordinates, the free energy of the incommensurate phases
up to the fourth order in the magnetic order parameters is
\begin{widetext}
\begin{align}
\label{eq:free_energy_IC}
f&=\big(aT-D-\lambda^{}_{+}(\qv^{}_{IC})\big)\left|\sigma^{}_{x}(\qv^{}_{IC})\right|^2+3b\left|\sigma^{}_{x}(\qv^{}_{IC})\right|^4
+\big(aT-\lambda^{}_{+}(\qv^{}_{IC})\big)\left|\sigma^{}_{b}(\qv^{}_{IC})\right|^2+3b\left|\sigma^{}_{b}(\qv^{}_{IC})\right|^4\nonumber\\
&+2b\left|\sigma^{}_{x}(\qv^{}_{IC})\right|^2\left|\sigma^{}_{b}(\qv^{}_{IC})\right|^2\big[2+\cos(2\varphi^{}_x-2\varphi^{}_b)\big]
+V^{}_{cell}\sum_{\alpha=1}^3\frac{{P^{}_{\alpha}}^2}{2\chi^{0}_{E,\alpha}}+r\left|\sigma^{}_{x}(\qv^{}_{IC})\right|\left|\sigma^{}_{b}(\qv^{}_{IC})\right|\sin(\varphi^{}_{x}-\varphi^{}_{b})P^{}_{b}.
\end{align}
\end{widetext}
This expression is obtained by keeping the Fourier components
$\qv=\pm\qv^{}_{IC}$ in the total free energy
$f=f^{}_{mag}+f^{}_{el}+f^{}_{int}$. Minimizing with respect to
the polarization components, we find the induced polarization
\begin{align}
\label{eq:polarization}
P^{}_x&=P^{}_z=0, \nonumber\\
P^{}_b&=-\frac{\chi^{0}_{E,b}r}{V^{}_{cell}}\left|\sigma^{}_{x}(\qv^{}_{IC})\right|\left|\sigma^{}_{b}(\qv^{}_{IC})\right|\sin(\varphi^{}_{x}-\varphi^{}_{b}).
\end{align}
Inserting Eqs. \eqref{eq:polarization} into Eq.
\eqref{eq:free_energy_IC}, we get
\begin{widetext}
\begin{align}
\label{eq:free_energy_IC2}
f&=\big(aT-D-\lambda^{}_{+}(\qv^{}_{IC})\big)\left|\sigma^{}_{x}(\qv^{}_{IC})\right|^2+3b\left|\sigma^{}_{x}(\qv^{}_{IC})\right|^4+\big(aT-\lambda^{}_{+}(\qv^{}_{IC})\big)\left|\sigma^{}_{b}(\qv^{}_{IC})\right|^2+3b\left|\sigma^{}_{b}(\qv^{}_{IC})\right|^4\nonumber\\
&+2b\left|\sigma^{}_{x}(\qv^{}_{IC})\right|^2\left|\sigma^{}_{b}(\qv^{}_{IC})\right|^2\big[2+\cos(2\varphi^{}_x-2\varphi^{}_b)-2\gamma\sin^2(\varphi^{}_x-\varphi^{}_b)\big],
\end{align}
where $\gamma$ is a dimensionless parameter given by
\begin{align}
\label{eq:gamma}
\gamma=\frac{\chi^{0}_{E,b}r^2}{8V^{}_{cell}b}.
\end{align}
In order to minimize the free energy \eqref{eq:free_energy_IC2},
the phase difference $\varphi^{}_x-\varphi^{}_b$ should be
$\pm\pi/2$. In addition, we show below that $\gamma$ is of order
$10^{-5}$. Hence the last factor in the square brackets of Eq.
\eqref{eq:free_energy_IC2} will be neglected in the description of
the magnetic phase transitions. The minimization of the free
energy \eqref{eq:free_energy_IC2} with respect to
$\left|\sigma^{}_{x}(\qv^{}_{IC})\right|$ and
$\left|\sigma^{}_{b}(\qv^{}_{IC})\right|$ yields
\label{eq:Incommensurate order parameters}
\begin{align}
\label{eq:Incommensurate order parameters}&\left|\sigma^{0}_{x}(\qv^{}_{IC})\right|=\sqrt{\frac{a\big(T^{(0)}_{N3}-T\big)}{6b}} \quad,\quad \left|\sigma^{0}_{b}(\qv^{}_{IC})\right|=0 \quad,\quad \text{$T^{(0)}_{N2}<T<T^{(0)}_{N3}$}, \nonumber\\
&\left|\sigma^{0}_{x}(\qv^{}_{IC})\right|=\sqrt{\frac{a\big(4T^{(0)}_{N3}-T^{(0)}_{N2}-3T\big)}{24b}}
\quad,\quad
\left|\sigma^{0}_{b}(\qv^{}_{IC})\right|=\sqrt{\frac{a\big(T^{(0)}_{N2}-T\big)}{8b}}
\quad,\quad \text{$T<T^{(0)}_{N2}$},
\end{align}
and the corresponding free energies are
\begin{align}
&f^{(0)}_{AF3}=-\frac{a^2\big(T^{(0)}_{N3}-T\big)^2}{12b} \quad,\quad \text{$T^{(0)}_{N2}<T<T^{(0)}_{N3}$}, \nonumber\\
&f^{(0)}_{AF2}=-\frac{a^2\big[4\big(T^{(0)}_{N3}-T\big)^2+\frac{8}{3}\big(T-T^{(0)}_{N3}\big)\big(T^{(0)}_{N3}-T^{(0)}_{N2}\big)+\frac{4}{3}\big(T^{(0)}_{N3}-T^{(0)}_{N2}\big)^2\big]}{32b}
\quad,\quad \text{$T<T^{(0)}_{N2}$},
\end{align}
\end{widetext}
with the transition temperature $T^{(0)}_{N2}$ given by
\begin{align}
\label{eq:T_N2} T^{(0)}_{N2}=T^{(0)}_{N3}-\frac{3D}{2a}.
\end{align}
By calculating the phase $\phi(\qv^{}_{IC})$ of
$J(\qv^{}_{IC};\tauv^{}_{1},\tauv^{}_{2})$ we can find the
magnetic structure of the phases AF3 and AF2. Using the
experimental incommensurate wave vector
$\qv^{}_{IC}=(-0.214,0.5,0.457)$, this phase is found to be
$\phi(\qv^{}_{IC})=2\pi y$ for the two sets of exchange couplings.
Using this relation and $\varphi^{}_x-\varphi^{}_b=\pm\pi/2$ in
Eqs. \eqref{eq:Fourier transform} and \eqref{eq:eigenvalues}, the
spins of the two Mn$^{2+}$ ions in the AF3 and AF2 phases are
\begin{align}
\label{eq:magnetic moments AF3, AF2}
&\;\Sv(\Rv+\tauv^{}_1)=\sqrt{2}\left|\sigma^{0}_{x}(\qv^{}_{IC})\right|\cos\big(\qv^{}_{IC}\cdot\Rv+\psi\big)\hat{\bf x}\nonumber\\
&\mp\sqrt{2}\left|\sigma^{0}_{b}(\qv^{}_{IC})\right|\sin\big(\qv^{}_{IC}\cdot\Rv+\psi\big)\hat{\bf b},\\
&\;\Sv(\Rv+\tauv^{}_2)=-\sqrt{2}\left|\sigma^{0}_{x}(\qv^{}_{IC})\right|\cos\big(\qv^{}_{IC}\cdot\Rv+\psi+\Delta\phi\big)\hat{\bf x}\nonumber\\
&\pm\sqrt{2}\left|\sigma^{0}_{b}(\qv^{}_{IC})\right|\sin\big(\qv^{}_{IC}\cdot\Rv+\psi+\Delta\phi\big)\hat{\bf
b}.
\end{align}
Here $\psi$ is an arbitrary phase and
$\Delta\phi\equiv\qv^{}_{IC}\cdot\left(\tauv^{}_{2}-\tauv^{}_{1}\right)+\phi(\qv^{}_{IC})-\pi=\pi
q^{}_{IC,c}$, with $q^{}_{IC,c}$ being the $c$ component of
$\qv^{}_{IC}$. Using the experimental value
$q^{}_{IC,c}=0.457$,\cite{LG93} this phase is
$\Delta\phi=0.457\pi$. This is exactly the magnetic structure
observed in neutron scattering studies.\cite{LG93} We emphasize
that while group theoretical analysis yields several magnetic
structures consistent with the crystal symmetries, the magnetic
structure described by Eqs. \eqref{eq:magnetic moments AF3, AF2}
is the actual structure observed in experiments. The two possible
signs correspond to the phase difference
$\varphi^{}_x-\varphi^{}_b=\pm\pi/2$ and represent spirals with
opposite chirality,
\begin{align}
&\;\Sv(\Rv+\tauv^{}_{1})\times\Sv(\Rv+\tauv^{}_{2})=\nonumber\\
&\pm2\left|\sigma^{0}_{x}(\qv^{}_{IC})\right|\left|\sigma^{0}_{b}(\qv^{}_{IC})\right|\sin\big(\Delta\phi\big)\hat{\bf
z}.
\end{align}
Here $\hat{\bf z}$ is a unit vector perpendicular to the spiral
plane. Various studies reveal that the spin chirality is strongly
correlated with the electric polarization and can be controlled by
poling the polarization with an external electric
field.\cite{SH08,FT10} This observation is in agreement with the
form \eqref{eq:polarization} of the electric polarization, in
which $\varphi^{}_x-\varphi^{}_b$ changes sign together with
$\Pv$.

Taking into account the magnetoelectric coupling in the
description of the magnetic phase transitions will introduce small
corrections to the transition temperature $T^{(0)}_{N2}$ and to
the order parameters in the AF2 phase. As mentioned above, these
corrections are governed by the dimensionless parameter $\gamma$
[see Eq. \eqref{eq:gamma}]. Using these corrections to the first
order in $\gamma$, we find that the electric susceptibility takes
the form
\begin{align}
\label{eq:electric susceptibility} \chi^{}_{E,b}(T)=
\begin{cases}
\chi^{0}_{E,b} & T>T^{(0)}_{N3}
\\
\chi^{0}_{E,b}\big(1+\frac{\widetilde{T}^{(0)}_{N2}-T^{(0)}_{N2}}{T-\tilde{T}^{(0)}_{N2}}\big)
& \widetilde{T}^{(0)}_{N2}<T<T^{(0)}_{N3}
\\
\chi^{0}_{E,b}\big(1+g(T)\frac{\widetilde{T}^{(0)}_{N2}-T^{(0)}_{N2}}{\tilde{T}^{(0)}_{N2}-T}\big)
& T^{(0)}_{N1}<T<\widetilde{T}^{(0)}_{N2}
\\
\chi^{0}_{E,b} & T<T^{(0)}_{N1},
\end{cases}
\end{align}
where $\widetilde{T}^{(0)}_{N2}$ is the shifted transition
temperature:
\begin{align}
\label{eq:Shifted_T_N2} \widetilde{T}^{(0)}_{N2}\approx
T^{(0)}_{N2}+\gamma\left(T^{(0)}_{N3}-T^{(0)}_{N2}\right).
\end{align}
The function $g(T)$ is
\begin{align}
\label{eq:g(T)}
g(T)=\frac{f^{}_2(T)}{f^{}_1(T)(T^{(0)}_{N3}-T^{(0)}_{N2})}-1,
\end{align}
where
\begin{align}
f^{}_1(T)&=-8T+\frac{32}{3}T^{(0)}_{N3}-\frac{8}{3}T^{(0)}_{N2}, \nonumber\\
f^{}_2(T)&=14T^2+\nu^{}_{1}T+\nu^{}_{2},
\end{align}
with
$\nu^{}_{1}=-\frac{1}{6}\left(95T^{(0)}_{N3}+73T^{(0)}_{N2}\right)$
and
$\nu^{}_{2}=16\left(T^{(0)}_{N3}\right)^2-\frac{97}{6}T^{(0)}_{N2}T^{(0)}_{N3}+\frac{85}{6}\left(T^{(0)}_{N2}\right)^2$.

The first order phase transition AF2$\rightarrow$AF1 can be
treated in the following way. Since the AF1 phase is characterized
by the commensurate wave vectors
$\qv^{}_{C1,2}=(\pm\frac{1}{4},\frac{1}{2},\frac{1}{2})$, we
calculate the free energy $f^{(0)}_{AF1}$ for this phase and then
look for a temperature below which $f^{(0)}_{AF1}<f^{(0)}_{AF2}$.
Since $\qv^{}_{C2}=-\qv^{}_{C1}+(0,1,1)$ we need to consider only
the Fourier components
$\qv=\pm\qv^{}_{C}=\pm(\frac{1}{4},\frac{1}{2},\frac{1}{2})$ in
Eq. \eqref{eq:magnetic_free_energy}. After some algebra we find
the free energy
\begin{align}
f&=\big(aT-D-\lambda^{}_{+}(\qv^{}_{C})\big)\left|\sigma^{}_{x}(\qv^{}_{C})\right|^2\nonumber\\
&+b\left|\sigma^{}_{x}(\qv^{}_{C})\right|^4\big[3+\cos(4\varphi-4\pi
y)\big].
\end{align}
Here $\varphi$ is the phase of $\sigma^{}_{x}(\qv^{}_{C})$,
determined to be $\pi(y+\frac{1}{4})$ in order to minimize the
free energy. Therefore the equilibrium order parameter and the
corresponding free energy are
\begin{align}
\label{eq:Commensurate order parameter}
\left|\sigma^{}_{x}(\qv^{}_{C})\right|=\sqrt{\frac{\lambda^{}_{+}(\qv^{}_{C})+D-aT}{4b}},
\end{align}
\begin{align}
f^{(0)}_{AF1}=-\frac{\big(\lambda^{}_{+}(\qv^{}_{C})+D-aT\big)^2}{8b}.
\end{align}
For the commensurate wave vector
$\qv^{}_{C}=(\frac{1}{4},\frac{1}{2},\frac{1}{2})$ we find the
phase $\phi(\qv^{}_{C})=2\pi y-\pi$ of
$J(\qv^{}_{C};\tauv^{}_{1},\tauv^{}_{2})$ for both sets of
exchange couplings. Using this relation and
$\varphi=\pi(y+\frac{1}{4})$ in Eqs. \eqref{eq:Fourier transform}
and \eqref{eq:eigenvalues}, the spins of the two Mn$^{2+}$ ions in
the AF1 phase are
\begin{align}
\label{eq:magnetic moments AF1}
&\Sv(\Rv+\tauv^{}_1)=\sqrt{2}\left|\sigma^{0}_{x}(\qv^{}_{C})\right|\cos\big(\qv^{}_{C}\cdot\Rv+\frac{\pi}{4}\big)\hat{\bf x}, \nonumber\\
&\Sv(\Rv+\tauv^{}_2)=-\sqrt{2}\left|\sigma^{0}_{x}(\qv^{}_{C})\right|\cos\big(\qv^{}_{C}\cdot\Rv-\frac{\pi}{4}\big)\hat{\bf
x}.
\end{align}
Equations \eqref{eq:magnetic moments AF1} describe a magnetic
structure of the type $\uparrow\uparrow\downarrow\downarrow$ along
both the $a$ and $c$ axes, in agreement with the structure
observed in experiments.\cite{LG93} We note again that this is the
observed structure out of the two possible structures suggested by
group theory.

The solution of the inequality $f^{(0)}_{AF1}<f^{(0)}_{AF2}$ is of
the form $T<T^{(0)}_{N1}$ provided that
\begin{align}
\label{eq:condition_T_N1}
\epsilon>\max\bigg\{2\left(1-\eta\right),\;\frac{2}{3}\left(1-\sqrt{3\eta^2-2}\right)\bigg\},
\end{align}
where $\epsilon\equiv\frac{D}{\lambda^{}_{+}(\qv^{}_{IC})}$ and
$\eta\equiv\frac{\lambda^{}_{+}(\qv^{}_{C})}{\lambda^{}_{+}(\qv^{}_{IC})}$.
In this case, the transition temperature $T^{(0)}_{N1}$ is given
by
\begin{align}
\label{eq:T_N1}
T^{(0)}_{N1}=\left[\frac{4\left(\eta^2-1\right)+4\epsilon-3\epsilon^2}{4\left(2\left(\eta-1\right)+\epsilon\right)}+\epsilon\right]\frac{\lambda^{}_{+}(\qv^{}_{IC})}{a}.
\end{align}
We study below the effects of magnetic field on the transition
temperatures.

\subsection{The effect of an external magnetic field}
The formalism presented above can be generalized to take into
account the effect of a uniform external magnetic field $\hv$.
This can be accomplished by adding to the free energy the Zeeman
term
$F^{}_{Z}=g\mu^{}_B\sum_{\Rv}\sum_{\tauv=\tauv^{}_{1},\tauv^{}_{2}}\Sv(\Rv+\tauv)\cdot\hv$,
or, equivalently\cite{comment2}
\begin{align}
\label{eq:Zeeman}
f^{}_{Z}\equiv\frac{F^{}_{Z}}{N}=g\mu^{}_B\sum_{\tauv=\tauv^{}_{1},\tauv^{}_{2}}\Sv(0,\tauv)\cdot\hv.
\end{align}
Minimizing the free energy with respect to
$S^{}_{\alpha}(0,\tauv)$ at the paramagnetic phase, we find the
response to the external magnetic field
\begin{align}
\label{Ferromagnetic Fourier component1}
S^{}_{\alpha}(0,\tauv)=-\frac{\chi^{}_{\alpha}(T)}{g
\mu^{}_B}h^{}_{\alpha} \qquad (\alpha=x,b),
\end{align}
with the magnetic susceptibility following a Curie-Weiss law
\begin{align}
\label{eq:magnetic_susceptibility}
\chi^{}_{\alpha}(T)=\frac{(g\mu^{}_B)^2}{aT-D^{}_{\alpha}-2\sum_{i=1}^{11}J^{}_i}.
\end{align}
Comparing Eq. \eqref{eq:magnetic_susceptibility} with the general
Curie-Weiss law\cite{Ashcroft&Mermin}
\begin{align}
\label{eq:magnetic_susceptibility2}
\chi^{}_{\alpha}(T)=\frac{\frac{(g\mu^{}_B)^2J(J+1)}{3k^{}_{B}}}{T-\theta^{}_{\alpha}},
\end{align}
we identify the parameter $a$ introduced in the expansion of the
entropy [see Eq. \eqref{eq:Entropy}] as
\begin{align}
\label{eq:a parameter} a=\frac{3k^{}_{B}}{J(J+1)}.
\end{align}
For Mn$^{2+}$ ions with $J=S=5/2$ this parameter is
$a^{}_{\text{Mn}}=0.343k^{}_{B}$. The Curie-Weiss temperature is
related to the exchange couplings and the anisotropy energy by:
\begin{align}
\label{eq:Curie-Weiss temperature}
\theta^{}_{\alpha}=\frac{J(J+1)}{3k^{}_{B}}\left(D^{}_{\alpha}+2\sum_{i=1}^{11}J^{}_i\right).
\end{align}

In the incommensurate phases AF3 and AF2, Eq. \eqref{Ferromagnetic
Fourier component1} is replaced by
\begin{align}
\label{Ferromagnetic Fourier component2}
S^{}_{\alpha}(0,\tauv)=\frac{-\chi^{}_{\alpha}(T)h^{}_{\alpha}}{g
\mu^{}_B\left[1+\frac{d^{}_{1\alpha}\left|\sigma^{0}_{x}(\qv^{}_{IC})\right|^2+d^{}_{2\alpha}\left|\sigma^{0}_{b}(\qv^{}_{IC})\right|^2}{a\left(T-\theta^{}_{\alpha}\right)}\right]},
\end{align}
where $d^{}_{1x}=d^{}_{2b}=12b$ and $d^{}_{2x}=d^{}_{1b}=4b$. The
corresponding form in the AF1 phase is
\begin{align}
\label{Ferromagnetic Fourier component3}
S^{}_{\alpha}(0,\tauv)=\frac{-\chi^{}_{\alpha}(T)h^{}_{\alpha}}{g
\mu^{}_B\left[1+\frac{e^{}_{\alpha}\left|\sigma^{0}_{x}(\qv^{}_{C})\right|^2}{a\left(T-\theta^{}_{\alpha}\right)}\right]},
\end{align}
with $e^{}_{x}=12b$ and $e^{}_{b}=4b$. The ferromagnetic Fourier
component at $\qv=0$ couples to the incommensurate and
commensurate wave vectors through the fourth order term in Eq.
\eqref{eq:magnetic_free_energy}. This coupling modifies the
coefficients of the free energy expansion and, consequently, the
transition temperatures. In the presence of an external magnetic
field, the first two transition temperatures are (to second order
in the magnetic field)
\begin{align}
\label{eq:T_3 with magnetic field}
\begin{cases}
T^{}_{N3}(h^{}_{x})=T^{(0)}_{N3}\left[1-12\frac{b\chi^{2}_{x}(T^{(0)}_{N3})}{aT^{(0)}_{N3}(g\mu^{}_B)^2}h^{2}_{x}\right] \quad & \hv=h^{}_{x}\hat{\bf x}, \\
T^{}_{N3}(h^{}_{b})=T^{(0)}_{N3}\left[1-4\frac{b\chi^{2}_{b}(T^{(0)}_{N3})}{aT^{(0)}_{N3}(g\mu^{}_B)^2}h^{2}_{b}\right]
\quad & \hv=h^{}_{b}\hat{\bf b},
\end{cases}
\end{align}
\begin{align}
\label{eq:T_2 with magnetic field}
\begin{cases}
T^{}_{N2}(h^{}_{x})=T^{(0)}_{N2} \quad & \hv=h^{}_{x}\hat{\bf x},\\
T^{}_{N2}(h^{}_{b})=T^{(0)}_{N2}\left[1-16\kappa\frac{b\chi^{2}_{b}(T^{(0)}_{N2})}{a(g\mu^{}_B)^2}h^{2}_{b}\right]
\quad & \hv=h^{}_{b}\hat{\bf b},
\end{cases}
\end{align}
with
$\kappa=\left(\frac{1}{T^{(0)}_{N2}}+\frac{8}{3\left(T^{(0)}_{N2}-\theta^{}_{b}\right)}\right)$.
For an external magnetic field along the easy axis direction, the
inequality which determines the stability range of the AF1 phase
is
\begin{widetext}
\begin{align}
\label{eq:T_1 with magnetic field_x}
T&<T^{(0)}_{N1}+\frac{8b}{2\left(\lambda^{}_{+}(\qv^{}_{C})-\lambda^{}_{+}(\qv^{}_{IC})\right)+D}\bigg\{T-3\frac{\lambda^{}_{+}(\qv^{}_{C})+D}{a}+2T^{(0)}_{N3}\nonumber\\
&\quad+\frac{1}{T-\theta^{}_{x}}\left[18\left(T-\frac{\lambda^{}_{+}(\qv^{}_{C})+D}{a}\right)^2-8\left(T^{(0)}_{N3}-T\right)^2\right]\bigg\}\left(\frac{\chi^{}_{x}(T)}{g\mu^{}_B}\right)^2h^{2}_{x},
\end{align}
while for a magnetic field along the $b$ direction it is
\begin{align}
\label{eq:T_1 with magnetic field_b}
T&<T^{(0)}_{N1}-\frac{8b}{2\left(\lambda^{}_{+}(\qv^{}_{C})-\lambda^{}_{+}(\qv^{}_{IC})\right)+D}\bigg\{T+\frac{\lambda^{}_{+}(\qv^{}_{C})+D}{a}-\frac{2}{3}\left(2T^{(0)}_{N2}+2T^{(0)}_{N2}\right)\nonumber\\
&\quad-\frac{1}{T-\theta^{}_{b}}\left[2\left(T-\frac{\lambda^{}_{+}(\qv^{}_{C})+D}{a}\right)^2-8\left(\frac{T^{(0)}_{N3}+2T^{(0)}_{N2}}{3}-T\right)^2\right]\bigg\}\left(\frac{\chi^{}_{x}(T)}{g\mu^{}_B}\right)^2h^{2}_{b}.
\end{align}
\end{widetext}
Equations \eqref{eq:T_3 with magnetic field}-\eqref{eq:T_1 with
magnetic field_b} describe the $T-H$ phase diagrams up to second
order in $h$.
\subsection{The effect of doping}
We can gain insight on the effect of small concentrations of
magnetic Fe$^{2+}$ or non-magnetic Zn$^{2+}$ and Mg$^{2+}$ ions at
the Mn sites in the following way. Assuming that the orbital
angular momentum is quenched, we set $J=S=2$ in Eq. \eqref{eq:a
parameter} and identify the parameter $a$ [see Eq.
\eqref{eq:Entropy}] for the Fe$^{2+}$ ion as
$a^{}_{\text{Fe}}=0.5k^{}_{B}$. Using this value, we get
\begin{equation}
\label{eq:Fe a} a(x)=a^{}_{\text{Mn}}x+a^{}_{\text{Fe}}(1-x),
\end{equation}
where $x$ is the Fe concentration. Since the exchange couplings of
Fe-Fe and Fe-Mn pairs as well as the anisotropy energy for the Fe
ion are not known, we assume a linear dependence of the quantities
$\lambda^{}_{+}(\qv^{}_{IC})$, $D$ and
$\eta=\frac{\lambda^{}_{+}(\qv^{}_{C})}{\lambda^{}_{+}(\qv^{}_{IC})}$
for small values of $x$:
\begin{align}
\label{eq:Fe parameters}
&\lambda^{}_{+}(\qv^{}_{IC}(x),x)=\lambda^{}_{+,IC}(0)+c^{}_1x, \nonumber\\
&D(x)=D(0)+c^{}_2x, \nonumber\\
&\eta(x)=\eta(0)+c^{}_3x.
\end{align}
We use the relations \eqref{eq:Fe parameters} in order to modify
the expressions \eqref{eq:T_N3}, \eqref{eq:T_N2} and
\eqref{eq:T_N1} for the transition temperatures. Then, by
expanding these expressions to first order in $x$ and fitting to
the slopes measured in experiments,\cite{CRP09} we are able to
extract the values of $c^{}_1$, $c^{}_2$ and $c^{}_3$. We neglect
any changes in the parameter $b$.

For the case of the non-magnetic Zn$^{2+}$ ion we set
$a^{}_{\text{Zn}}=D^{}_{\text{Zn}}=0$ as well as
$J^{\text{Zn-Mn}}_{i}=J^{\text{Zn-Zn}}_{i}=0$, and find the
$x$-dependence of the different parameters
\begin{align}
\label{eq:Zn parameters}
&\lambda^{}_{+}(\qv,x)=\lambda^{}_{+}(\qv)(1-x)^2, \nonumber\\
&a(x)=a(0)(1-x), \nonumber\\
&D(x)=D(0)(1-x), \nonumber\\
&\eta(x)=\eta(0).
\end{align}
Using these relations the first two transition temperatures are
given by
\begin{align}
\label{eq:non-magnetic}
&T^{}_{N3}(x)=T^{(0)}_{N3}-\frac{\lambda^{}_{+}(\qv^{}_{IC})}{a}x, \nonumber\\
&T^{}_{N2}(x)=T^{(0)}_{N2}-\frac{\lambda^{}_{+}(\qv^{}_{IC})}{a}x.
\end{align}
These results explain the linear decrease of $T^{}_{N3}$ and of
$T^{}_{N2}$ as a function of $x$ observed in
experiments.\cite{CRP11,ML09} The treatment of the
AF2$\rightarrow$AF1 transition is much more subtle and will be
discussed below. We note that all the results above do not depend
on the type of the non-magnetic ion. This is in agreement with the
observed similarities of the transition temperatures in Zn$^{2+}$
and Mg$^{2+}$ doping.\cite{CRP11,ML09}
\section{Comparison with experiments}
\label{Sec 3} In this section we compare the results of the
preceding section with different experimental observations and
examine the consistency of the phase diagrams with the
experimental sets of exchange couplings of Ehrenberg \textit{et
al.} and Ye \textit{et al.}. The results of the preceding section
can be used to fit the parameters of the model within the Landau
theory. We use Eqs. \eqref{eq:T_N3} and \eqref{eq:T_N2} with
$a^{}_{\text{Mn}}=0.343k^{}_{B}$ [see Eq. \eqref{eq:a parameter}]
and the experimental transition temperatures $T^{(0)}_{N3}$ and
$T^{(0)}_{N2}$ in order to extract the values of the parameters
$\lambda^{}_{+}(\qv^{}_{IC})$ and $D$ for MnWO$^{}_{4}$. Using the
experimental values $T^{(0)}_{N3}\approx13.5K$ and
$T^{(0)}_{N2}\approx12.3-12.7K$, these parameters are found to be
$\lambda^{}_{+}(\qv^{}_{IC})\approx4.36-4.45K$ and
$D=0.27-0.18k^{}_{B}K$. The ratio
$\eta=\frac{\lambda^{}_{+}(\qv^{}_{C})}{\lambda^{}_{+}(\qv^{}_{IC})}$
is then chosen to be $\eta\approx0.97-0.98$ in order to fit Eq.
\eqref{eq:T_N1} to the experimental transition temperature
$T^{(0)}_{N1}\approx7-8K$. These values are consistent with the
condition \eqref{eq:condition_T_N1}.

Next we use Eqs. \eqref{eq:magnetic_susceptibility2} and
\eqref{eq:T_3 with magnetic field}-\eqref{eq:T_1 with magnetic
field_b}, with the experimental Curie-Weiss temperature
$\theta^{}_{x}\approx\theta^{}_{b}\approx-75K$,\cite{AAH06,DH69}
and calculate the $T-H$ phase diagrams by fitting the parameter
$b$. In order to get the best fit to the experimental phase
diagram of Arkenbout \textit{et al.},\cite{AAH06} the parameter
$b$ was chosen to be $0.095k^{}_{B}K$. Figure \ref{fig:T H Phase
Diagrams} shows the results. The calculated and the experimental
phase diagrams are in good agreement. Discrepancies at low
temperatures or at high fields are expected due to the finite
expansion of the free energy, which is terminated at fourth order.
\begin{figure}[ht]
\begin{center}
\subfigure[]{ \label{fig:H||x}
\includegraphics[width=0.45\textwidth]{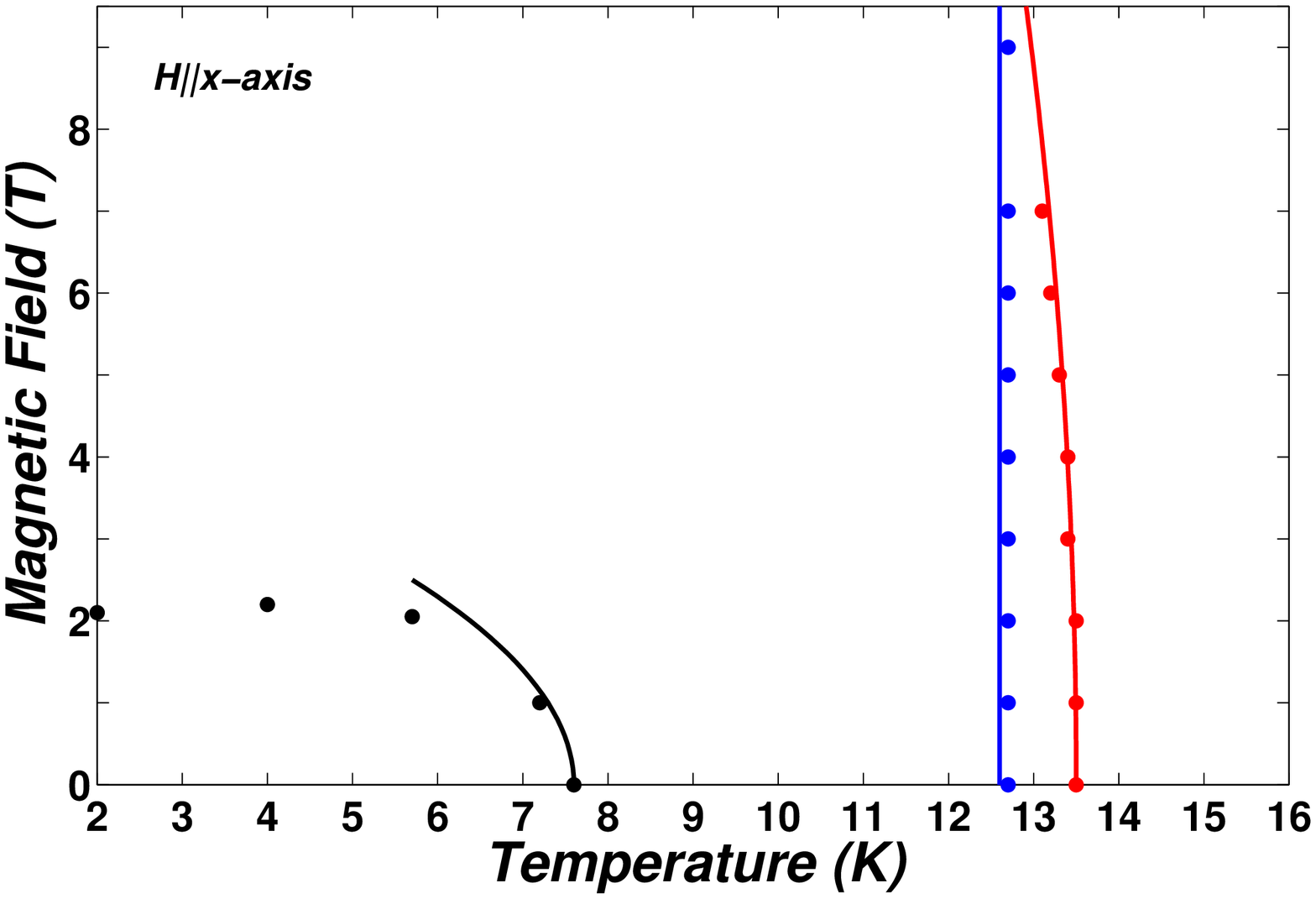}
} \subfigure[]{ \label{fig:H||b}
\includegraphics[width=0.45\textwidth]{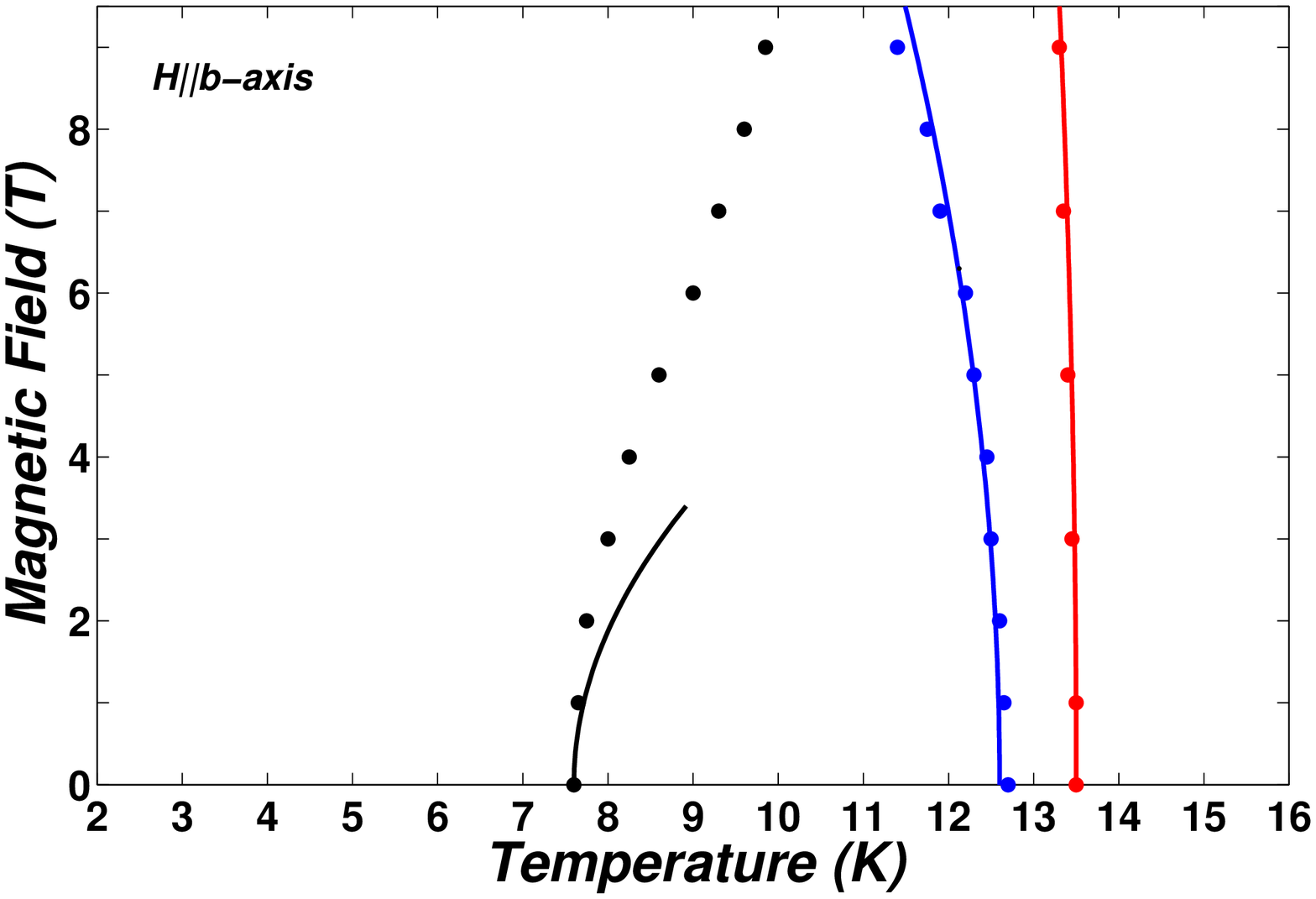}
}
\end{center}
\caption{\label{fig:T H Phase Diagrams}(Color online)
Magnetoelectric phase diagrams of MnWO$^{}_4$ with magnetic fields
parallel to the (a) easy and (b) b axes. The solid lines are the
calculated transition temperatures and the dots describe the
experimental points of Arkenbout \textit{et al.}.\cite{AAH06} The
calculated phase diagrams were obtained by setting
$b=0.095k^{}_{B}K$.}
\end{figure}

\begin{figure}[ht]
\centering
\includegraphics[width=0.45\textwidth]{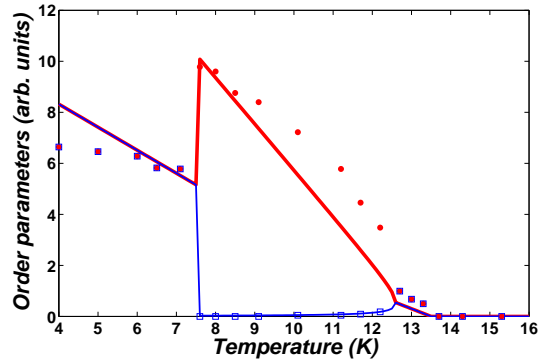}
\caption{\label{fig:Order Parameters}(Color online) The
temperature dependence of
$\left(\left|\sigma^{0}_{x}(\qv^{}_{IC})\right|\pm\left|\sigma^{0}_{b}(\qv^{}_{IC})\right|\right)^2+\left|\sigma^{0}_{x}(\qv^{}_{C})\right|^2$.
The red (thick) line corresponds to the + sign and the blue (thin)
one to the minus sign. The integrated intensities of the
polarized-neutron diffraction (scaled by 0.002) from Ref.
\onlinecite{SH08} are the red dots and the blue empty squares.}
\end{figure}

The development of the magnetic order parameters with decreasing
temperature has been studied by polarized-neutron
diffractions.\cite{SH08} Generally, the magnetic moment at site
$\tauv$ belonging to the unit cell at the lattice point $\Rv$ can
be written as
\begin{align}
\mv(\Rv+\tauv)&=m^{}_{x}\cos\left(\qv\cdot\Rv+\phi^{}_{\tau}\right)\hat{\bf x}\nonumber\\
&+m^{}_{b}\sin\left(\qv\cdot\Rv+\phi^{}_{\tau}\right)\hat{\bf b}.
\end{align}
The cross-sections for polarized-neutron scattering, where the
neutrons are polarized parallel and anti-parallel to the
scattering vector, are given by\cite{SH08}
\begin{align}
I=I{}_{0}\left(m^{}_{x}\pm m^{}_{b}\right)^2,
\end{align}
with $I^{}_{0}$ being a constant. Using Eqs. \eqref{eq:magnetic
moments AF3, AF2} and \eqref{eq:magnetic moments AF1}, we see that
these cross-sections are proportional to
$\left(\left|\sigma^{0}_{x}(\qv^{}_{IC})\right|\pm\left|\sigma^{0}_{b}(\qv^{}_{IC})\right|\right)^2+\left|\sigma^{0}_{x}(\qv^{}_{C})\right|^2$.
Then, from the second of Eqs. \eqref{eq:Incommensurate order
parameters}, the magnetic order parameters in the AF2 phase can be
written as
\begin{align}
\label{eq:Incommensurate order parameters3}\left|\sigma^{0}_{x}(\qv^{}_{IC})\right|&=\sqrt{\frac{a\left[\frac{4}{3}\left(T^{(0)}_{N3}-T^{(0)}_{N2}\right)+T^{(0)}_{N2}-T\right]}{8b}}, \nonumber\\
\left|\sigma^{0}_{b}(\qv^{}_{IC})\right|&=\sqrt{\frac{a\big(T^{(0)}_{N2}-T\big)}{8b}}.
\end{align}
Tol{\'e}dano \textit{et al.}\cite{TP10} assumed that
$\left|\sigma^{0}_{x}(\qv^{}_{IC})\right|$ is fixed below
$T^{(0)}_{N2}$. According to the first of Eqs.
\eqref{eq:Incommensurate order parameters3}, such an assumption is
valid only for
$T^{(0)}_{N2}-T\ll\frac{4}{3}\left(T^{(0)}_{N3}-T^{(0)}_{N2}\right)$.
At lower temperatures this assumption is inconsistent with the
evolution of the observed integrated intensities reported in Ref.
\onlinecite{SH08}, which show that both
$\left|\sigma^{0}_{x}(\qv^{}_{IC})\right|$ and
$\left|\sigma^{0}_{b}(\qv^{}_{IC})\right|$ continue to grow below
$T^{}_{N2}$, with the ellipticity
$p\equiv\frac{m^{}_{b}}{m^{}_{x}}=\frac{\left|\sigma^{0}_{b}(\qv^{}_{IC})\right|}{\left|\sigma^{0}_{x}(\qv^{}_{IC})\right|}$
approaching 1 (so that the spiral is almost circular) as the
temperature decreases. Therefore, we preferred to use the explicit
dependence of $\left|\sigma^{0}_{x}(\qv^{}_{IC})\right|$ on the
temperature. Using Eqs. \eqref{eq:Incommensurate order
parameters3}, the ellipticity below $T^{}_{N2}$ can be written as
\begin{align}
p=\frac{1}{\sqrt{1+\omega}},
\end{align}
where
$\omega\equiv\frac{4\left(T^{0}_{N3}-T^{0}_{N2}\right)}{3\left(T^{0}_{N2}-T\right)}$.
Since the difference $T^{0}_{N3}-T^{0}_{N2}\approx0.8K$ is very
small in the case of MnWO$^{}_{4}$, the ellipticity rapidly
approaches 1 with decreasing temperature in the spiral phase AF2.
The small difference $T^{0}_{N3}-T^{0}_{N2}$ for MnWO$^{}_{4}$ is
a consequence of the small single-ion anisotropy of Mn$^{2+}$
ions. This should be compared with the case of TbMnO$^{}_{3}$, for
which $T^{}_{N3}\approx42K$ and $T^{}_{N2}\approx27K$. In this
multiferroic, the ellipticity grows much more slowly with
decreasing temperature,\cite{YY07} due to the large difference
$T^{0}_{N3}-T^{0}_{N2}\approx15K$, which is in turn a result of
the larger single-ion anisotropy of Mn$^{3+}$ ions. In Fig.
\ref{fig:Order Parameters} we sketch the quantities
$\left(\left|\sigma^{0}_{x}(\qv^{}_{IC})\right|\pm\left|\sigma^{0}_{b}(\qv^{}_{IC})\right|\right)^2+\left|\sigma^{0}_{x}(\qv^{}_{C})\right|^2$
from Eqs. \eqref{eq:Incommensurate order parameters3} and
\eqref{eq:Commensurate order parameter} together with the
experimental data points of Ref. \onlinecite{SH08}.

\begin{figure}[ht]
\centering
\includegraphics[width=0.45\textwidth]{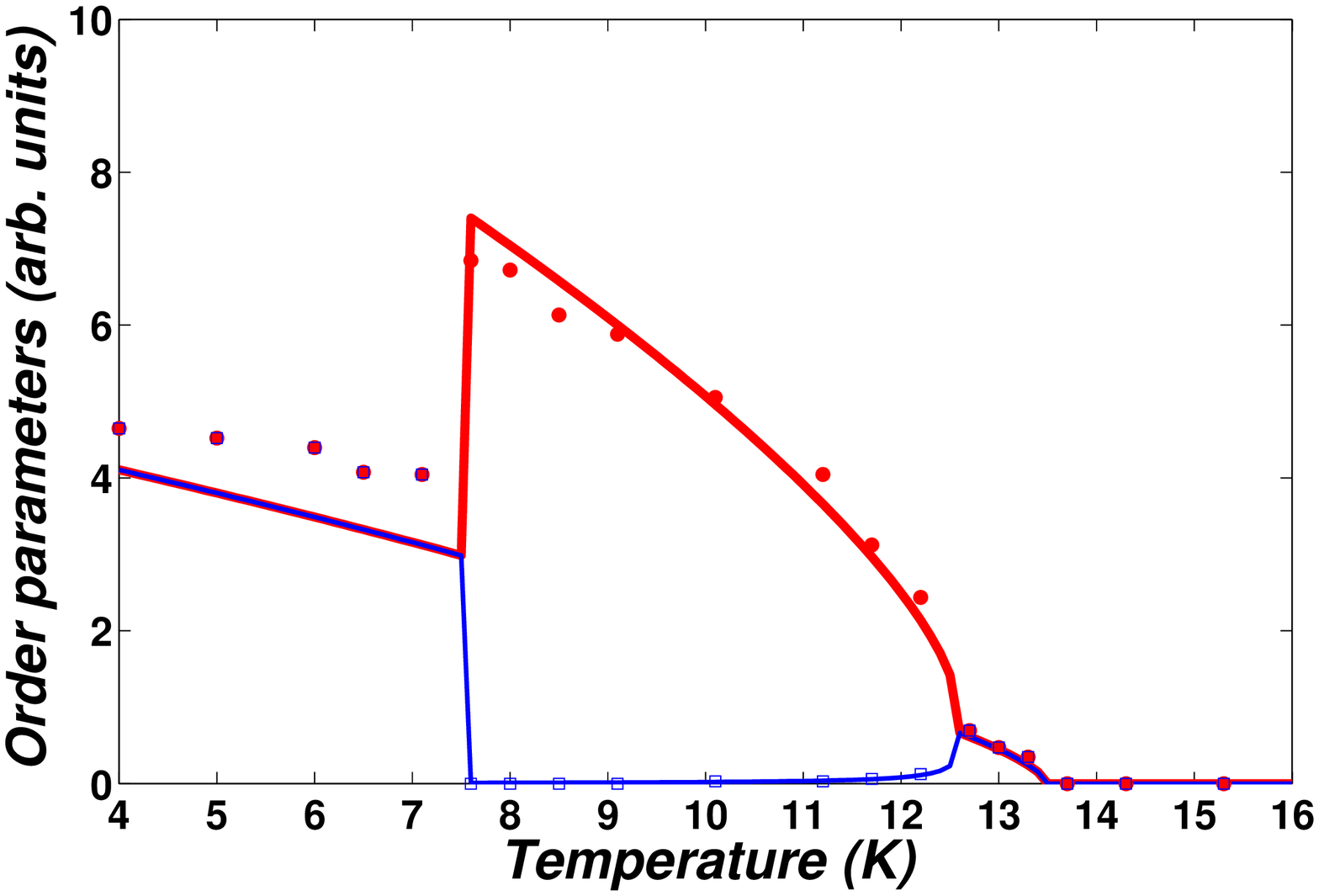}
\caption{\label{fig:Order Parameters_Critical}(Color online) The
temperature dependence of
$\left(\left|\sigma^{0}_{x}(\qv^{}_{IC})\right|\pm\left|\sigma^{0}_{b}(\qv^{}_{IC})\right|\right)^2+\left|\sigma^{0}_{x}(\qv^{}_{C})\right|^2$
with the critical exponent $\beta\approx1/3$. The red (thick) line
corresponds to the + sign and the blue (thin) one to the minus
sign. The integrated intensities of the polarized-neutron
diffraction (scaled by 0.0014) from Ref. \onlinecite{SH08} are the
red dots and the blue empty squares.}
\end{figure}

The development of the calculated order parameters is in a
qualitative agreement with the temperature dependence of the
integrated intensities. However, for
$T^{(0)}_{N2}-T\gg\frac{4}{3}\left(T^{(0)}_{N3}-T^{(0)}_{N2}\right)$
in the AF2 phase, the quantity
$\left(\left|\sigma^{0}_{x}(\qv^{}_{IC})\right|+\left|\sigma^{0}_{b}(\qv^{}_{IC})\right|\right)^2$
is linear in $T$, in contradiction with the temperature dependence
of the integrated intensity, as can be seen in Fig. \ref{fig:Order
Parameters}. A possible explanation for this apparent discrepancy
is related to fluctuations near the transitions, that are not
taken into account by the mean-field Landau
theory.\cite{Landau&Lifshitz} As pointed out in Ref.
\onlinecite{HAB08}, the transition P$\rightarrow$AF3 belongs to
the universality class of the XY model, while the transition
AF3$\rightarrow$AF2 belongs to the Ising universality class. Hence
we present in Fig. \ref{fig:Order Parameters_Critical} the same
quantities as in Fig. \ref{fig:Order Parameters}, but replacing
the square roots of Eqs. \eqref{eq:Incommensurate order
parameters3} and \eqref{eq:Commensurate order parameter} by the
critical exponent $\beta=1/3$, roughly appropriate for these two
models. As seen from the figure, these revised expressions are in
good agreement with the observed integrated intensities. This
behavior illustrates the possible importance of fluctuations in
MnWO$^{}_{4}$. Further consequences of fluctuations near the
transitions will be discussed below in the context of the Ginzburg
criterion.

\begin{figure}[ht]
\centering
\includegraphics[width=0.45\textwidth]{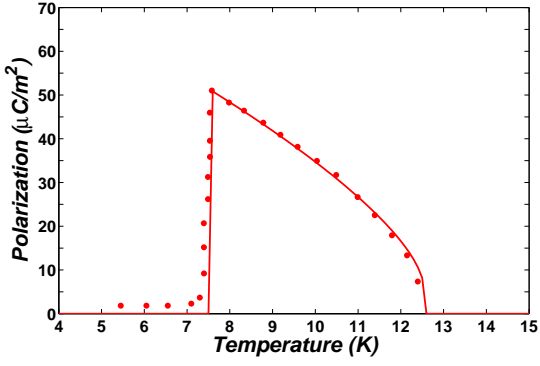}
\caption{\label{fig:polarization_critical}(Color online) The
ferroelectric polarization calculated with the critical exponent
$\beta\approx\frac{1}{3}$. The solid lines are the calculated
quantities and the dots are the experimental points of Taniguchi
\textit{et al.}.\cite{TK06} The calculated polarization was
obtained by setting
$\frac{\chi^{0}_{E,b}\left|r\right|}{V^{}_{cell}}=27.5\mu C/m^2$.}
\end{figure}

The magnetoelectric coupling $r$ is determined by fitting Eq.
\eqref{eq:polarization} to the experimental data of the induced
ferroelectric polarization.\cite{TK06} The ferroelectric
polarization is plotted in Fig. \ref{fig:polarization}. The best
fit to the experimental data is obtained for the value
$\frac{\chi^{0}_{E,b}\left|r\right|}{V^{}_{cell}}=21\mu C/m^2$. In
addition, the electric susceptibility for $T>T^{(0)}_{N3}$ (in the
paraelectric and paramagnetic phase), is experimentally found to
be $\chi^{0}_{E,b}=11.3\epsilon^{}_0$.\cite{TK06} The
dimensionless parameter $\gamma$ [see Eq. \eqref{eq:gamma}] is
then $\gamma=5.9\cdot10^{-5}$. This value supports the assumption
that the magnetic transitions are almost unaffected by the
magnetoelectric coupling. The dielectric constant
$\epsilon^{}_b=1+\frac{\chi^{0}_{E,b}}{\epsilon^{}_0}$ is shown in
Fig. \ref{fig:susceptibility}. This result is in good agreement
with the experimental measurements of Ref. \onlinecite{TK06}. The
narrow width of the divergence region is a consequence of the
small difference between $\widetilde{T}^{(0)}_{N2}$ and
${T}^{(0)}_{N2}$.
\begin{figure}[ht]
\begin{center}
\subfigure[\label{fig:polarization}]{ \label{fig:polarization}
\includegraphics[width=0.45\textwidth]{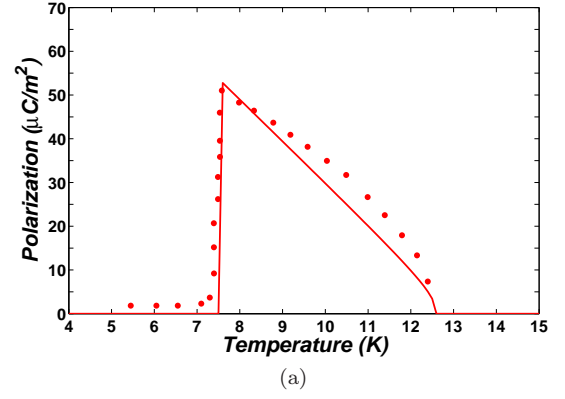}
} \subfigure[\label{fig:susceptibility}]{
\label{fig:susceptibility}
\includegraphics[width=0.45\textwidth]{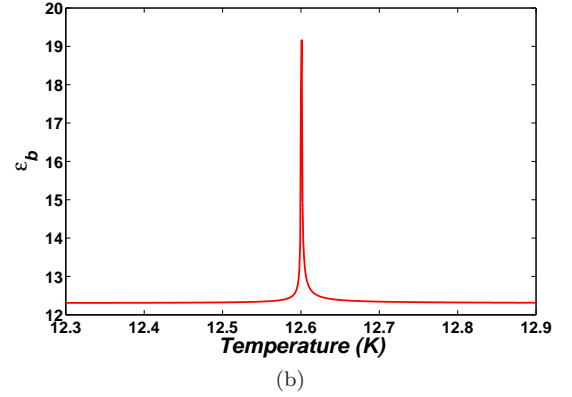}
}
\end{center}
\caption{\label{fig:electric}(Color online) (a) The ferroelectric
polarization and (b) the dielectric constant $\epsilon^{}_b$. The
solid lines are the calculated quantities and the dots are the
data points of Taniguchi \textit{et al.}.\cite{TK06} The
calculated polarization was obtained by setting
$\frac{\chi^{0}_{E,b}\left|r\right|}{V^{}_{cell}}=21\mu C/m^2$.}
\end{figure}
Once again, the discrepancy between the linear behavior of the
calculated polarization and the observed one may be reconciled by
assuming a critical exponent $\beta\approx\frac{1}{3}$ for the
magnetic order parameters. The behavior of the calculated
polarization in this case is given in Fig.
\ref{fig:polarization_critical}.

To examine the effect of Fe doping, we use the relations
\eqref{eq:Fe a} and \eqref{eq:Fe parameters} in the expressions
for the transition temperatures and fit the slope to the
experimental value according to the $x-T$ phase diagram of
Chaudhury \textit{et al.}.\cite{CRP09} This procedure yields the
values $c^{}_1\approx-3.26k^{}_{B}K$,
$c^{}_2\approx13.03k^{}_{B}K$ and $c^{}_3\approx-1.3$. The
anisotropy energy increases with increasing Fe concentration, as
expected, since as opposed to the Mn$^{2+}$ ion, the Fe$^{2+}$ ion
possesses a non-vanishing angular momentum.\cite{HN10}

\begin{figure}[ht]
\centering
\includegraphics[width=0.45\textwidth]{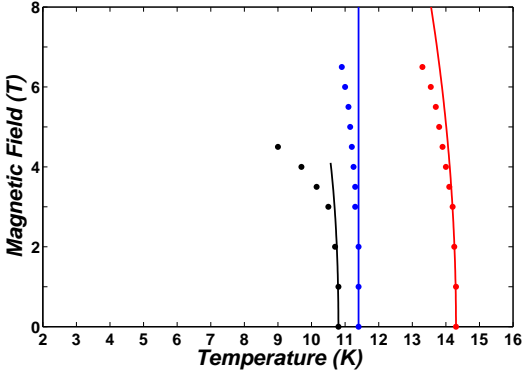}
\caption{\label{fig:Fe}(Color online) Magnetoelectric phase
diagram of Mn$^{}_{0.965}$Fe$^{}_{0.035}$WO$_4$ with the magnetic
field parallel to the easy axis. The solid lines are the
calculated transition temperatures and the dots are the
experimental points of Ye \textit{et al.}.\cite{YF08}}
\end{figure}

Calculating the different parameters for a small Fe concentration
$x$ and repeating the calculations of the $T-H$ phase diagram, we
can check the consistency of the above results. The resulting
phase diagram for $x=0.035$ is shown in Fig. \ref{fig:Fe}. Except
for high fields or low temperatures, the result is in fine
agreement with the measurement of Ye \textit{et al.}.\cite{YF08}
The reentrant ferroelectric phase observed at low
temperatures\cite{CRP09,CRP09b} may be explained by higher order
terms in the free energy expansion.

\begin{figure}[ht]
\centering
\includegraphics[width=0.45\textwidth]{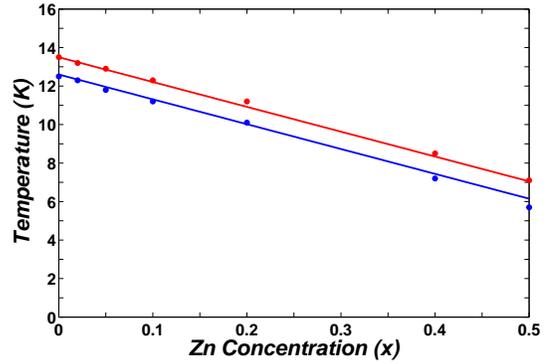}
\caption{\label{fig:Zn}(Color online) Transition temperatures
$T^{}_{N3}(x)$ and $T^{}_{N2}(x)$ of
Mn$^{}_{1-x}$Zn$^{}_{x}$WO$^{}_4$. The solid lines are the
calculated transition temperatures and the dots are the data
points of Chaudhury \textit{et al.}.\cite{CRP11}}
\end{figure}

The effect of non-magnetic ions on the transition temperatures
$T^{}_{N3}(x)$ and $T^{}_{N2}(x)$ is given by Eq.
\eqref{eq:non-magnetic}. These results are drawn in Fig.
\ref{fig:Zn} together with the experimental data of Chaudhury
\textit{et al.}\cite{CRP11} of Mn$^{}_{1-x}$Zn$^{}_{x}$WO$^{}_4$.
Similar results have been observed in
Mn$^{}_{1-x}$Mg$^{}_{x}$WO$^{}_4$.\cite{ML09} We stress that
unlike the case of Fe doping, the results for the transition
temperatures $T^{}_{N3}(x)$ and $T^{}_{N2}(x)$ in the case of
non-magnetic ions doping do not require additional
phenomenological parameters.

As opposed to $T^{}_{N3}(x)$ and $T^{}_{N2}(x)$, the calculated
transition temperature $T^{}_{N1}(x)$ does not coincide with the
experimentally measured one.\cite{CRP11} The discrepancy may be
explained by allowing small changes in the exchange couplings
$J^{\text{Mn-Mn}}_{i}$ due to spin-lattice coupling (or exchange
striction). In other words, if we assume that
$J^{\text{Mn-Mn}}_{i}(x)=J^{\text{Mn-Mn}}_{i}(1+\xi^{}_{i}x)$ with
$\xi^{}_{i}x\ll1$, then $T^{}_{N1}(x)$ changes dramatically while
$T^{}_{N3}(x)$ and $T^{}_{N2}(x)$ are almost not influenced. The
reason for this behavior is that the transition temperature
$T^{}_{N1}$ [see Eq. \eqref{eq:T_N1}] is much more sensitive to
small changes in the exchange couplings than the transition
temperatures $T^{}_{N3}$ and $T^{}_{N2}$ [see Eqs. \eqref{eq:T_N3}
and \eqref{eq:T_N2}]. A significant spin-lattice coupling in the
multiferroic MnWO$^{}_4$ has been demonstrated \cite{TK08} by the
appearance of an incommensurate lattice modulation in the AF3 and
AF2 phases, with a lattice propagation vector equal to twice the
magnetic propagation vector. In addition, thermal expansion
measurements reveal considerable discontinuities in the lattice
parameters at the AF2$\rightarrow$AF1 first order phase
transition.\cite{CRP08b} Another indication for a dependence of
the Mn-Mn exchange couplings on the non-magnetic dopant
concentration is provided by the small change of the
incommensurate propagation vector from
$\qv^{}_{IC}=(-0.214,0.5,0.457)$ in MnWO$^{}_4$ to
$\qv^{}_{IC}=(-0.209,0.5,0.453)$ in
Mn$^{}_{0.85}$Zn$^{}_{0.15}$WO$^{}_4$.\cite{ML09}

The next step is to compare the above fitted parameters with the
parameters calculated directly from the experimental sets of
exchange couplings of Ehrenberg \textit{et al.} and Ye \textit{et
al.}. The calculated exchange couplings of Ref. \onlinecite{TC09}
yield much higher transition temperatures than the observed ones
and thus will not be discussed here. Indeed, The problem of
overestimation of exchange interactions by DFT calculations has
been indicated by the authors.\cite{TC09}

The first step is to maximize $\lambda^{}_{+}(\qv)$ [see Eq.
\eqref{eq:lambda}] in order to find the incommensurate wave vector
$\qv^{}_{IC}$ and the corresponding eigenvalue
$\lambda^{}_{+}(\qv^{}_{IC})$. The maximization process yields
$\qv^{}_{IC}=(-0.28,0.5,0.44)$ and
$\lambda^{}_{+}(\qv^{}_{IC})=3.82k^{}_{B}K$ for the
$J^{}_1-J^{}_9$ values of Ehrenberg \textit{et al.},\cite{EH99}
while for the $J^{}_1-J^{}_{11}$ values of Ye \textit{et
al.}\cite{YF11} we find $\qv^{}_{IC}=(-0.3,0.5,0.49)$ and
$\lambda^{}_{+}(\qv^{}_{IC})=3.85k^{}_{B}K$. These results are in
qualitative agreement with the incommensurate wave vector
$\qv^{}_{IC}=(-0.214,\frac{1}{2},0.457)$ observed in experiments.
However, the differences are not negligible, suggesting possible
errors in the experimental sets of exchange couplings. In
addition, the transition temperatures $T^{(0)}_{N3}$ and
$T^{(0)}_{N2}$ calculated from Eqs. \eqref{eq:T_N3} and
\eqref{eq:T_N2} with $a^{}_{\text{Mn}}=0.343k^{}_{B}$ [see Eq.
\eqref{eq:a parameter}] are found to be $T^{(0)}_{N3}=12.79K$,
$T^{(0)}_{N2}=10.3K$ for the set of Ehrenberg \textit{et al.} and
$T^{(0)}_{N3}=13.67K$, $T^{(0)}_{N2}=10K$ for the set of Ye
\textit{et al.}. These values slightly differ from the observed
transition temperatures, especially the second one. The ratio
$\eta=\frac{\lambda^{}_{+}(\qv^{}_{C})}{\lambda^{}_{+}(\qv^{}_{IC})}$
is found to be $\eta=0.974$ and $\eta=0.97$ for the sets of
Ehrenberg \textit{et al.} and Ye \textit{et al.}, respectively.
Table ~\ref{tab:Parameters_ summary} summarizes the values of
$\lambda^{}_{+}(\qv^{}_{IC})$, $D$ and $\eta$ calculated from the
experimental sets of magnetic parameters and those fitted to the
experimental transition temperatures.
\begin{table}[ht]
\caption{\label{tab:Parameters_ summary} Comparison between the
model parameters calculated from the experimental sets of Ref.
\onlinecite{EH99} and Ref. \onlinecite{YF11} and those fitted to
the experimental transition temperatures.}
    \begin{tabular}{c|c|c|c}
    \hline
    \hline
    Parameter & Ref. \onlinecite{EH99} & Ref. \onlinecite{YF11} & This work \\ \hline
    $\lambda^{}_{+}(\qv^{}_{IC})(k^{}_{B}K)$ & 3.82 & 3.85 & 4.36-4.45 \\
    &  &  \\ \hline
    $D(k^{}_{B}K)$ & 0.568 & 0.83 & 0.27-0.18 \\
    &  &  \\ \hline
    $\eta$ & 0.974 & 0.97 & 0.97-0.98 \\\hline \hline
    \end{tabular}
\end{table}

The calculation of the Curie-Weiss temperature reveals a much more
serious discrepancy. According to Eq. \eqref{eq:Curie-Weiss
temperature}, the Curie-Weiss temperature is $\theta^{}_x=-7.6K$,
$\theta^{}_b=-9.25K$ for the set of Ehrenberg \textit{et al.} and
$\theta^{}_x=-23.2K$, $\theta^{}_b=-25.65K$ for the set of Ye
\textit{et al.}. These values do not fit the experimental
Curie-Weiss temperature $\theta\approx-75K$.\cite{AAH06,DH69} We
suspect that the origin of most of the discrepancies are errors in
the set of magnetic couplings. The results suggested by our model
may be used as additional constraints in the determination of
those couplings. As mentioned before, an additional possible cause
for the above discrepancies is related to fluctuations near the
transitions, as will be discussed in the next section.
\section{The Ginzburg criterion}
\label{Sec 4} The results of the preceding sections have been
obtained within the mean-field approximation. Here we estimate the
Ginzburg range, in which fluctuations become important, near the
first transition P$\rightarrow$AF3, by two methods. First we
compare the mean square fluctuation of the order parameter
$\sigma^{}_{x}(\qv^{}_{IC})$ with the mean-field value, and then
we compare the discontinuity in the heat capacity derived from the
Landau theory with the divergent heat capacity, originating from
the fluctuations at quadratic order.\cite{Landau&Lifshitz}

Let us denote by
$\delta\sigma^{}_{x}(\qv)=\sigma^{}_{x}(\qv)-\langle\sigma^{}_{x}(\qv)\rangle$
the fluctuation of the order parameter in the AF3 phase. The
correlation function of these deviations is
\begin{align}
\label{eq:Correlation}
\langle\delta\sigma^{}_{x}(\qv)\delta\sigma^{}_{x}(\qv')\rangle=\frac{k^{}_{B}T\delta^{}_{\qv',-\qv}}{4N\left(D+\lambda^{}_{+}(\qv)-aT\right)},
\end{align}
where $N$ is the number of unit cells in the correlation volume.
We can find the correlation lengths by expanding
$\lambda^{}_{+}(\qv)$ to second order around $\qv^{}_{IC}$:
\begin{align}
\label{eq:second order expansion} \lambda^{}_{+}(\qv)\approx
\lambda^{}_{+}(\qv^{}_{IC})+\sum_{i,j}M^{}_{ij}\left(q^{}_{i}-q^{}_{IC,i}\right)\left(q^{}_{j}-q^{}_{IC,j}\right),
\end{align}
with $M^{}_{ij}\equiv\frac{1}{2}\frac{\partial^2
\lambda^{}_{+}(\qv)}{\partial q^{}_{i}\partial
q^{}_{j}}\bigg|^{}_{\qv=\qv^{}_{IC}}$. Denoting by $\mu^{}_1$,
$\mu^{}_2$ and $\mu^{}_3$ the three eigenvalues of the positive
matrix $-M^{}_{ij}$, the three correlation lengths are
\begin{align}
\label{eq:correlation lengths}
\xi^{}_{i}=\sqrt{\frac{\mu^{}_{i}}{a\left(T^{(0)}_{N3}-T\right)}}.
\end{align}
Substituting $\qv=\qv^{}_{IC}$ and
$N=\frac{\xi^{}_{1}\xi^{}_{2}\xi^{}_{3}}{V^{}_{cell}}$ in Eq.
\eqref{eq:Correlation}, the condition
$\langle\left|\delta\sigma^{}_{x}(\qv^{}_{IC})\right|^2\rangle\ll\left|\sigma^{0}_{x}(\qv^{}_{IC})\right|^2$
for the validity of the mean-field theory
reads\cite{Landau&Lifshitz}
\begin{align}
\label{eq:Ginzburg}
\frac{k^{}_{B}T^{(0)}_{N3}}{4a\left(T^{(0)}_{N3}-T\right)}\frac{V^{}_{cell}}{\xi^{}_{1}\xi^{}_{2}\xi^{}_{3}}\ll
\frac{a\left(T^{(0)}_{N3}-T\right)}{6b}.
\end{align}
Inserting Eq. \eqref{eq:correlation lengths} into Eq.
\eqref{eq:Ginzburg} at the Ginzburg temperature $T^{}_G$, we find
\begin{align}
\label{eq:Ginzburg temperature1}
\left|T^{}_{G}-T^{(0)}_{N3}\right|\approx
\frac{9k^{2}_{B}b^{2}V^{2}_{cell}\left(T^{(0)}_{N3}\right)^2}{4a\mu^{}_{1}\mu^{}_{2}\mu^{}_{3}}.
\end{align}
Equation \eqref{eq:Ginzburg temperature1} estimates the
temperature range below $T^{(0)}_{N3}$, in which fluctuations are
not negligible.

Let us now estimate the Ginzburg range according to the second
method. On the one hand, according to Landau theory, the heat
capacity $c=-T\frac{\partial^2 f}{\partial T^2}$ grows
discontinuously at the transition P$\rightarrow$AF3:
\begin{align}
\label{eq:heat capacity Landau} \Delta c^{}_{L}\equiv
c^{}_{L}\left(T^{(0)-}_{N3}\right)-c^{}_{L}\left(T^{(0)+}_{N3}\right)=\frac{a^2T^{(0)}_{N3}}{6b}.
\end{align}
On the other hand, assuming fluctuations at quadratic order, the
singular part of the heat capacity is given by
\begin{align}
\label{eq:heat capacity Fluctuations}
c^{}_{G}=\frac{V^{}_{cell}k^{}_{B}a^{2}T^{2}}{2\left(2\pi\right)^3}\int_{BZ}\frac{d^3q}{\left(aT-D-\lambda^{}_{+}(\qv)\right)^2},
\end{align}
where the integral is over the first Brillouin zone. In the
neighborhood of $T^{(0)}_{N3}$, the main contribution to the
integral comes from the neighborhood of the incommensurate wave
vector $\qv^{}_{IC}$ in reciprocal space. Thus we can use the
expansion \eqref{eq:second order expansion}. Replacing the first
Brillouin zone by a sphere, and taking $T\approx T^{(0)}_{N3}$, we
can estimate the integral in Eq. \eqref{eq:heat capacity
Fluctuations}:
\begin{align}
\label{eq:heat capacity Fluctuations2} c^{}_{G}\approx
\frac{k^{}_{B}a^{1.5}T^{2}\left(T-T^{(0)}_{N3}\right)^{-0.5}}{16\pi\sqrt{\mu^{}_{1}\mu^{}_{2}\mu^{}_{3}}}.
\end{align}
Comparing Eqs. \eqref{eq:heat capacity Landau} and \eqref{eq:heat
capacity Fluctuations2} at the Ginzburg temperature $T^{}_G$, we
find\cite{Landau&Lifshitz}
\begin{align}
\label{eq:Ginzburg temperature2}
\left|T^{}_{G}-T^{(0)}_{N3}\right|\approx
\left(\frac{6}{16\pi}\right)^2\frac{k^{2}_{B}b^{2}V^{2}_{cell}\left(T^{(0)}_{N3}\right)^2}{a\mu^{}_{1}\mu^{}_{2}\mu^{}_{3}}.
\end{align}
Calculating the eigenvalues $\mu^{}_1$, $\mu^{}_2$ and $\mu^{}_3$
from the experimental sets of exchange couplings, the Ginzburg
temperature is estimated to be
$\left|T^{}_{G}-T^{(0)}_{N3}\right|\approx 9.41K$ and
$\left|T^{}_{G}-T^{(0)}_{N3}\right|\approx 6.24K$ for the sets of
Ehrenberg \textit{et al.} and Ye \textit{et al.}, respectively, by
the first method [see Eq. \eqref{eq:Ginzburg temperature1}] while
it is $\left|T^{}_{G}-T^{(0)}_{N3}\right|\approx 0.06K$ and
$\left|T^{}_{G}-T^{(0)}_{N3}\right|\approx 0.04K$ by the second
method [see Eq. \eqref{eq:Ginzburg temperature2}]. These values
suggest that fluctuations of the order parameters can also
contribute to the discrepancies between the experimental data and
the mean-field Landau theory results.
\section{Summary and Conclusions}
\label{Sec 5} We have studied the phase diagram of
Mn$^{}_{1-x}$M$^{}_{x}$WO$^{}_{4}$ (M=Fe, Zn, Mg) by a
semi-phenomenological Landau theory. The energy has been modelled
by a Heisenberg Hamiltonian with a single-ion anisotropy, while
the entropy has been expanded in powers of the classical spins.
This approach is different from the previous theoretical
studies,\cite{TP10,SVP10} which are purely phenomenological, since
it enables to compare different sets of exchange couplings.
Although a purely phenomenological approach may capture all the
symmetry aspects of the problem and may provide a full mapping of
the stable states allowed by the order parameter
symmetries,\cite{TP10} it does not indicate a clear connection
between the free energy coefficients and the microscopic
interactions. The advantage of our approach is the simple relation
of the free energy coefficients with experimentally derived
quantities such as the superexchange couplings and the anisotropy
coefficients. For instance, this simple relation allows us to
consider the effect of different dopants on the phase diagram, not
discussed in Ref. \onlinecite{TP10}. We emphasize that our
approach does not contradict any symmetry requirement.

We used the superexchange interaction couplings from the inelastic
neutron scattering studies of Ehrenberg \textit{et al.}\cite{EH99}
and Ye \textit{et al.}.\cite{YF11} The results show that both sets
yield transition temperatures $T^{(0)}_{N3}$ and $T^{(0)}_{N2}$
that slightly deviate from the experimental temperatures, and
significantly underestimate the Curie-Weiss temperature
$\left|\theta\right|$. In addition, the calculated incommensurate
wave vector $\qv^{}_{IC}$ has non-negligible deviations from the
experimentally observed one. The results presented here can serve
as additional constraints on a future determination of the
magnetic Hamiltonian parameters. Another possible cause for the
discrepancies relates to fluctuations near the transitions. We
have demonstrated the possible important contribution of
fluctuations in MnWO$^{}_{4}$. This issue should be further
examined in future experiments.

Beyond that, the model clarifies the effect of different dopants
on the phase diagram. The sensitivity of the expression
\eqref{eq:T_N1} for the transition temperature $T^{}_{N1}(x)$ to
small changes of the ratio
$\eta\equiv\frac{\lambda^{}_{+}(\qv^{}_{C})}{\lambda^{}_{+}(\qv^{}_{IC})}$
reflects the frustrated nature of the multiferroic MnWO$^{}_4$.
The origin of the complex phase diagram lies in the competition
between different superexchange interactions. Small changes in the
local environment of the Mn$^{2+}$ ions due to a chemical doping
cause a significant change in the phase diagram. The sensitivity
for the local environment manifests itself by the contrasting
behavior of doping with different ions.

Looking to the future, two points should be further examined.
Firstly, a new analysis of the inelastic scattering experiments,
together with the additional constraints provided in this work,
should improve the exchange couplings for the multiferroic
MnWO$^{}_4$. Secondly, the measurement of the critical exponents
near the transitions would shed light on the effect of
fluctuations. This may contribute to the general understanding of
critical phenomena in multiferroics.
\begin{acknowledgments}
We thank H. Shaked for helpful discussions. We acknowledge support
from the Israel Science Foundation (ISF).
\end{acknowledgments}


\begin{references}
\bibitem{comment1} The values of the experimental parameters were adjusted to the definitions of Eq. \eqref{eq:Hamiltonian}, in which each pair of spins is counted once, and the anisotropy term includes a prefactor $\frac{1}{2}$.
\bibitem{comment2} We note that Eq.\eqref{eq:Zeeman} has a positive sign since the magnetic moment of electrons is opposite to the spin direction.
\end{references}
\end{document}